\documentclass{aa}  

\usepackage{graphicx}
\graphicspath{{figs/}}
\usepackage{txfonts}
\usepackage{ulem}
\usepackage[dvipsnames]{xcolor}
\usepackage{hyperref}
\hypersetup{
    colorlinks=true,       % false: boxed links; true: colored links
    linkcolor=blue,          % color of internal links
    citecolor=blue,        % color of links to bibliography
    filecolor=blue,      % color of file links
    urlcolor=blue           % color of external links
}
\usepackage[version=3]{mhchem}
\usepackage{orcidlink}
\definecolor{orcidlogocol}{rgb}{0.65, 0.807, 0.223}
\newcommand{\orcid}[1]{\orcidlink{#1}}

\usepackage{placeins}
\usepackage{siunitx}

\newcommand{\Tex}{T_\text{ex}}
\newcommand{\Tkin}{T_\text{kin}}
\newcommand{\Tbg}{T_\text{bg}}

\newcommand{\kB}{k_\text{B}}
\newcommand{\Ntot}{N_\text{tot}}
\newcommand{\NHtwo}{N_{\text{H}_2}}
\newcommand{\Aij}{A_{ij}}

\newcommand{\npH}{$n_{\text{p-H}_2}$}

\defcitealias{tanious2024}{Paper I}

\begin{document} 

   \title{Anatomy of the Class I protostar L1489 IRS with NOEMA}
   \subtitle{II. A disk replenished by a massive streamer} 

   \titlerunning{Anatomy of the Class I protostar L1489 IRS with NOEMA - II} 

   \author{M. Tanious\inst{1,2}\orcid{0009-0002-9761-8546}
          \and
          R. Le Gal\inst{1,2}\orcid{0000-0003-1837-3772}
          \and
          A. Faure\inst{1}\orcid{0000-0001-7199-2535}
          \and
          S. Maret \inst{1}\orcid{0000-0003-1104-4554}
          \and
          A. López-Sepulcre\inst{1,2}\orcid{0000-0002-6729-3640}
          \and
          P. Hily-Blant\inst{1}\orcid{0000-0003-3488-8442}
          }

    \institute{Université Grenoble Alpes, CNRS, IPAG, F-38000 Grenoble, France\\
            \email{
            [maxime.tanious ; romane.le-gal; alexandre.faure]@univ-grenoble-alpes.fr 
            }
        \and
            IRAM, 300 rue de la piscine, F-38406 Saint-Martin d’H\`{e}res, France
             }

   \date{Received 23 May 2025 / Accepted 16 September 2025}

% \abstract{}{}{}{}{} 
% 5 {} token are mandatory
 
  \abstract
    {
    Streamers are newly identified channels that transport mass from large, molecular-cloud scales down to small, protoplanetary-disk scales. To better understand their impact on planet formation, it is essential to study their physical and chemical properties.
    }
    {
    In this framework, we aim to characterize the longest streamer identified in carbon chain emission within the Class I system L1489 IRS, connecting the nearby prestellar core L1489 to the young stellar object (YSO).
    }   
   {
   We observed multiple transitions of \ce{C2H}, ortho-c-\ce{C3H2,} and \ce{HC3N} in L1489 IRS with NOEMA and IRAM-30m at 3mm and 2mm. Using a variety of radiative transfer methods, including a hyperfine structure (HFS) fitting,  rotational diagrams, and proposing a new self-consistent Markov chain Monte Carlo (MCMC) approach combined with the non-LTE \texttt{RADEX} code, we derived the column densities and abundances of those molecules, as well as the \ce{H2} number density along the streamer. This enabled us to estimate its mass, infall rate, and its impact on the \{star+disk\} system's mass.
   }
   {
   We found lower limits on the streamer mass of $\geq (4.67-18.3)\, \times 10^{-3}$~$M_\odot$ (i.e.,~$\geq 0.65-2.57$ times the current disk mass) and an infall rate of $\geq (1.94-7.57)\, \times 10^{-7}$~$M_\odot$~yr$^{-1}$, where the ranges correspond to the different molecular tracers. These values are consistent with those derived in similar Class I objects. 
   This suggests that the disk could be fully replenished by streamer material. Given its mass, the streamer is likely at the origin of the external warped disk seen in this system, as predicted by numerical simulations. Moreover, the first investigations based on the \ce{C2H}/c-\ce{C3H2} and \ce{HC3N}/c-\ce{C3H2} abundance ratios suggest that the streamer chemistry may be inherited from the core. These results suggest, for the first time, that the chemical composition of a Class I object is affected by a streamer connecting a Class I YSO to its natal environment.
   }
   {
   We demonstrate that the streamer in L1489 IRS has a significant impact on its disk. To better constrain how the streamer influences the disk's chemistry and determine whether its composition is inherited from the nearby core, further molecular surveys will be necessary toward the prestellar core, the streamer, and the YSO. Our findings reinforce the importance of characterizing the natal environment of protoplanetary disks both physically (e.g., structure formation) and chemically (e.g., material enrichment)  to fully understand their evolution.
   }


   \maketitle
%
%-------------------------------------------------------------------

\section{Introduction}

For many years, tail structures ($\sim10^3$~au) have been detected in the vicinity of protoplanetary disks. They are observed either in dust \citep[e.g.,][]{rebollido2024,zurlo2024} or in gas \citep[mainly through CO and its isotopologs; e.g.,][]{tang2012,huang2020,huang2021,huang2022,huang2023,kido2025}, or both \citep[e.g., in SU Aur; e.g.,][]{grady2001,akiyama2019,ginski2021}. They are likely associated to gravitational perturbations of the disk, either from internal (e.g., forming planets) or external agents \citep[e.g., accretion or flybys, e.g.,][]{thies2011,cuello-review2023}. Late infall might also create such structures and can significantly change disk morphologies, creating misalignments or warps \citep[e.g.,][]{thies2011,kuffmeier2021,kuffmeier2024}.

This late infall is usually seen as funneled streams (possibly in a more extended infalling envelope) and have therefore been referred to as ``streamers'' \citep{pineda2023}. Larger numbers of streamers continue to be identified, detected around young stellar objects (YSOs) at all evolutionary stages: at the very early Class 0 phase \citep[e.g.,][]{pineda2020, thieme2022}, in young and intermediate Class I objects \citep[e.g.,][]{valdivia-mena2022,hsieh2023,tanious2024}, and at the late Class II stage \citep[e.g.,][]{yen2019,akiyama2019,alves2020,garufi2021}. They  connect the large scales in molecular clouds to small scales in disks, thereby acting as a newly identified channel to deliver material to disk scales. Since their impact on disks is not yet fully understood, recent studies (e.g., NOEMA's Large Program PRODIGE; PIs: Caselli \& Henning) have focused on physical properties of streamers, characterizing their mass and infall rate, in particular. They may also have a significant impact on the disk chemistry \citep{podio2024}, but this has been poorly studied so far. In the context of better understanding  interstellar heritage, studying streamers impact on protoplanetary disks is crucial, both physically and chemically.

So far, streamers have been observed in multiple species: 
mainly in CO and isotopologs \citep{yen2014,akiyama2019,ginski2021,thieme2022,valdivia-mena2022,kido2023,gupta2023,tipsy2024}, 
but also in other O-bearing molecules \citep[e.g., \ce{HCO+} or \ce{H2CO},][]{yen2019, valdivia-mena2022, valdivia-mena2023, tipsy2024, podio2024, tanious2024}, 
N-bearing molecules \cite[e.g., \ce{HC3N}, \ce{CH3CN},][]{pineda2020, hsieh2023, valdivia-mena2023, valdivia-mena2024, tanious2024}, 
S-bearing molecules \citep[e.g., \ce{CS}, \ce{C2S}, \ce{SO}, \ce{SO2}, e.g.][]{garufi2022, codella2024, hanawa2024,taniguchi2024, tanious2024}, 
deuterated species \citep[e.g., \ce{DCN},][]{hsieh2023}, 
and carbon chains \citep[e.g., \ce{C2H}, \ce{c-C3H2},][]{taniguchi2024,tanious2024}. 
This offers a range of possibilities to study the physics, using molecular tracers probing different physical conditions, as well as the chemistry in these objects, looking, for instance, at gradients of molecular distributions along streamers and probe the C/N/O/S elemental ratios, or at the D/H isotopic ratio \citep{podio2024}, all known to have a crucial impact in shaping the composition of planet formation \citep{oberg2023}.

\object{L1489 IRS} (a.k.a. \object{IRAS 04016+2610}) is an ideal target to investigate the impact of streamers on protoplanetary disks. It is a Class I object located in the nearby Taurus star forming region \citep[$d\sim146$~pc,][]{roccatagliata2020} and, thus, it can be spatially resolved in a straightforward way using interferometers such as ALMA or NOEMA. It contains three nested disks, with the most external one Keplerian, warped, extending to 600~au \citep{sai2020,yamato2023}, and three confirmed streamers: two $\sim1400$~au-long seen in \ce{C^18O} emission \citep{yen2014} and a longer one ($\geq3000$~au) that was recently identified mainly on the basis of its carbon chain emission, for instance, \ce{C2H}, c-\ce{C3H2}, and \ce{HC3N} \citep[hereafter \citetalias{tanious2024}]{tanious2024}. The latter is likely at the origin of the external warped disk observed in this system and also likely connecting the YSO to the nearby prestellar core L1489 \citepalias{tanious2024}, which would still feed material to the YSO's disk \citep{brinch2007a}. The streamers likely impact the disk's midplane resulting in suggested accretion shocks seen in SO emission \citep[\citetalias{tanious2024}]{yen2014,yamato2023}. Those shocks are of particular interest as their physical conditions would release molecules trapped in the icy mantle of grains into the gas phase, likely affecting the chemistry in the disk.

Given those results, we aimed to pursue the analysis of the streamer started in \citetalias{tanious2024} to derive its properties, namely, its mass and infall rate, and start to investigate its chemistry. We describe in Sect.~\ref{sec:observations} the observations conducted and their data reduction. In Sect.~\ref{sec:methods}, we present the methods used to derive column densities and abundances of targeted molecules. This led to the results in Sect.~\ref{sec:results} that are discussed in Sect.~\ref{sec:discussion}. Finally, we summarize our main conclusions in Sect.~\ref{sec:conclusions}.

%%%%%%%%%%%%%%%%%%%%%%%%%%%%%%%%%%%%%%%%%%%%%%%%%%%%%%%%%%%

\section{Observations}
\label{sec:observations}

\subsection{NOEMA observations}

\begin{table*}[!ht]
\caption{Observed lines used in this work.}             
\label{table:lines}      
\centering                          
\renewcommand{\arraystretch}{1.5} 
\small
\begin{tabular}{l c c c c c c c c c}        
\hline\hline                 
    Species & Transition & $E_\text{up}$ & $g_\text{up}$ & $A_{ij}$ & Rest Freq. & $\theta_\text{maj} \times \theta_\text{min}$ (PA) & rms$_\text{chan}$ & $\Delta v$ & Telescope \\
    & & (K) & & (s$^{-1}$) & (GHz) & (\arcsec $\times$ \arcsec, \degr) & (mK) & (km~s$^{-1}$) & \\

\hline    
    \ce{C2H} & $1_{1.5,1}\to0_{0.5,0}$  & 4.191 & 3 & \num{1.272e-6} & 87.328585 & $2.12\times1.47$ (-157.6) & 177 & 0.21 & NOEMA + 30m \\
    
             & $1_{1.5,1}\to0_{0.5,1}$  & 4.191 & 3 & \num{2.599e-7} & 87.284105 & $2.12\times1.47$ (-157.6) & 185 & 0.21 & NOEMA + 30m \\
             
             & $1_{1.5,2}\to0_{0.5,1}$  & 4.193 & 5 & \num{1.531e-6} & 87.316898 & $2.12\times1.47$ (-157.6) & 179 & 0.21 & NOEMA + 30m \\
             
             & $1_{0.5,1}\to0_{0.5,0}$  & 4.197 & 3 & \num{2.613e-7} & 87.446470 & $2.11\times1.47$ (-157.6) & 168 & 0.21 & NOEMA + 30m \\
             
             & $1_{0.5,1}\to0_{0.5,1}$  & 4.197 & 3 & \num{1.275e-6} & 87.401989 & $2.11\times1.47$ (-157.6) & 169 & 0.21 & NOEMA + 30m \\
             
             & $1_{0.5,0}\to0_{0.5,1}$  & 4.197 & 1 & \num{1.536e-6} & 87.407165 & $2.11\times1.47$ (-157.6) & 169 & 0.21 & NOEMA + 30m \\

\hline  
    \ce{c}-\ce{C3H2}\tablefootmark{$\dagger$} & $2_{1,2}\to1_{0,1}$ & 6.445  & 15 & \num{2.322e-5} & 85.338894 & $2.22\times1.50$ (-159.6) & 32 & 7.03 & NOEMA + 30m \\
    
                           & $3_{1,2}\to2_{2,1}$ & 16.05 & 21 & \num{6.768e-5} & 145.089606 & $2.67\times2.36$ (116.0) & 183 & 0.52 & NOEMA + 30m \\
                           
                           & $3_{1,2}\to3_{0,3}$ & 16.05 & 21 & \num{9.919e-6} & 82.966200 & $2.23\times1.57$ (21.9) & 184 & 0.23 & NOEMA + 30m \\
                           
                           & $4_{1,4}\to3_{0,3}$ & 19.31 & 27 & \num{1.638e-4} & 150.851908 & $2.78\times2.33$ (44.7) & 241 & 0.50 & NOEMA + 30m \\

\hline 
    \ce{HC3N} & $8\to7$   & 15.72  & 17 & \num{2.941e-5} & 72.783822 & $35.63\times35.63$ (0)   & 90  & 0.20 & 30m \\
              & $9\to8$   & 19.65  & 19 & \num{4.215e-5} & 81.881468 & $31.67\times31.67$ (0)   & 47  & 0.18 & 30m \\
              & $10\to9$  & 24.02  & 21 & \num{5.812e-5} & 90.979023 & $28.50\times28.50$ (0)   & 192 & 0.16 & 30m \\
              & $11\to10$ & 28.82  & 23 & \num{7.769e-5} & 100.076392 & $1.82\times1.30$ (24.4) & 188 & 0.19 & NOEMA + 30m \\
              & $11\to10$ & 28.82  & 23 & \num{7.769e-5} & 100.076392 & $25.91\times25.91$ (0)  & 39  & 0.15 & 30m \\
              & $12\to11$ & 34.06  & 25 & \num{1.012e-4} & 109.173634 & $23.75\times23.75$ (0)  & 226 & 0.13 & 30m \\

\hline                                   
\end{tabular}
\tablefoot{The energies, degeneracies, Einstein coefficients and rest frequencies are taken from the CDMS \citep{muller2001,muller2005}. Critical densities of each lines are available in Table~\ref{table:critical-densities}. The per channel-rms is estimated as the mean value from line-free channels of cubes after primary beam correction. $\Delta v$ gives the velocity resolution for each cube.\\
\tablefoottext{$\dagger$}{All considered lines of \ce{c}-\ce{C3H2} are ortho lines.}
}
\end{table*}

We carried out 3mm interferometric observations of \object{L1489~IRS} with the NOrthern Extended Millimiter Array (NOEMA) in Band 1 (Project IDs: S20AH and W20AJ,  PI: Le Gal). They consist of single-field observations centered on \object{L1489~IRS} using the C- and A-array configurations between November 10, 2020 and March 21, 2021, yielding synthesized beam sizes $\sim 1\farcs4$ at 90~GHz using natural weighting. The PolyFiX correlator was configured in its high spectral resolution mode, enabling us to observe  multiple spectral windows at 62.5~kHz resolution simultaneously (i.e.,~$\sim0.2$~km~s$^{-1}$), along with a $\sim15.5$~GHz bandwidth with a resolution of 2~MHz. The sanity of interferometric data was checked by deconvolving $uv$ tables with the H\"ogbom algorithm \citep{hogbom1974}. We refer to \citetalias{tanious2024} for more details about these observations, their calibration, and reduction.

This dataset was completed with 2mm mosaic observations using NOEMA Band 2 receivers (Project ID: S24AR,  PIs: Tanious \& Le Gal). The mosaic is made of 52 individual pointings. The primary beam of the NOEMA antennas is $\sim$\,35\arcsec~at $\sim$\,145~GHz. The phase tracking center of the mosaic was located in $\alpha$(J2000)~=~04\textsuperscript{h}04\textsuperscript{m}47\fs257, $\delta$(J2000)~=~26\degr19\arcmin02\farcs884. 
The observations were carried out with the D-array configuration using 10 antennas, from July 23, 2024 to September 17, 2024, with projected baselines ranging from 15.1~m to 176~m. These observations yield a synthesized beam $\sim$\,2\farcs5~at 145~GHz using natural weighting. 
3C~84 served as the bandpass calibrator, LkH$\alpha$~101 served as flux calibrator, and QSO~B0400+258 and QSO~B0333+3208 served as the phase and amplitude calibrators. The uncertainty on derived intensities due to the calibration are below 10\%. 

Two frequency setups were observed using the NOEMA's correlator set in its survey mode, enabling to cover a total instantaneous bandwidth of $\sim$15.5~GHz per polarization and per setup with a spectral resolution of 250~kHz. The first spectral setup ranged in frequency from 134.195 to 142.322~GHz and 149.683 to 157.810~GHz, observed for 10~h on source; whereas the second one ranged in frequency from 141.695 to 149.823~GHz and 157.183 to 161.255~GHz, observed for 8.4~h on source. This resulted in a continuous bandwidth between 134.2 and 161.3~GHz at a $\sim$\,0.5 km~s$^{-1}$ velocity resolution. The data calibration was performed using the dedicated NOEMA pipeline in the \texttt{CLIC} software, which is part of the \texttt{GILDAS}\footnote{\url{http://www.iram.fr/IRAMFR/GILDAS}} distribution. Visibilities exceeding a phase rms of 61$\degr$ or a seeing of $1\farcs3$ were flagged as default parameters of the pipeline.

\subsection{IRAM-30m observations}
The 3mm short-spacings observations of \object{L1489~IRS} were carried with the IRAM-30m telescope (Project ID: 184-20, PI: Le Gal) between July 21 and 26, 2021, using the on-the-fly position-switching observing mode. We used the \textit{Eight MIxer Receiver} (EMIR) E090 connected to the narrow fast Fourier Transform Spectrometers (FTS~50) backends, which offer a bandwidth of $4 \time 1.82$~GHz by spectral setup at a 50~kHz resolution. After averaging the polarizations, a 30~MHz window around the line rest frequency was extracted, and a first-order baseline removed excluding line emission. More details about the observations and their reduction can be found in \citetalias{tanious2024}.

The 2mm short-spacings were observed on September 10, 2024 with the IRAM-30m telescope (Project ID: 097-24, PIs: Tanious \& Le Gal). We used EMIR E150 connected to the wide FTS~200 backends, which offer a bandwidth $\sim$\,16.2~GHz by spectral setup at a 200~kHz resolution. Two frequency setups were needed to cover the full NOEMA bandwidth. 
A 5.7\arcmin $\times$\,4.8\arcmin\, region centered on the NOEMA's mosaic phase tracking center was mapped using the on-the-fly position-switching observing mode. The half power primary beam of the IRAM-30m telescope is $\sim$\,17\arcsec~at $\sim$\,145~GHz. The reference position was set to $\alpha$(J2000)~=~04\textsuperscript{h}05\textsuperscript{m}09\fs214, $\delta$(J2000)~=~26\degr18\arcmin56\farcs390 (i.e., an offset of $\Delta\alpha=351\farcs5$, $\Delta\delta=0\farcs0$~from \object{L1489 IRS}). Single-pointing spectra acquired in frequency-switching mode toward the reference position showed no contamination from our targeted lines. Mars and 3C~84 were used as calibrators, as well as pointing and focus sources for the observations. The pointing and the focus were checked approximately every $\sim$\,1.5~h. The 2mm dataset was reduced in the same way as the 3mm dataset.

\subsection{Combination of IRAM-30m and NOEMA data}
We used the methods described in \citetalias{tanious2024} to combine IRAM-30m and NOEMA data \citep[i.e.,~using the pseudo-visibilities technique,][]{iram-memo-2008-2} and to deconvolve the 3mm dataset \citep[i.e.,~using the Multi Resolution Clean (MRC) algorithm of][down to twice the expected thermal noise with natural weighting and no mask]{mrc1988}. The data cubes were then corrected by the primary beam response down to 20\%.

The imaging of continuum-subtracted line cubes of the 2mm dataset was performed with the \texttt{IMAGER}\footnote{\url{https://imager.oasu.u-bordeaux.fr}} software, using the H\"ogbom deconvolution algorithm \citep{hogbom1974} with a natural weighting, down to twice the expected thermal noise. 
While the \texttt{IMAGER} documentation mentions applying the JvM correction \citep{jvm} as a possible adjustment for mosaics with varying uv-coverage in some cases, we first assessed the necessity of this correction by analyzing the synthesized beam variations between fields. Given the modest differences in beam areas (ratio < $10-20$\%) and the first-order nature of the JvM correction (c.f. \texttt{IMAGER} documentation Section 6.5.5), we opted not to apply it to avoid introducing unnecessary approximations.

\subsection{Maps descriptions}

\begin{figure}
    \centering
    \includegraphics[width=0.73\linewidth]{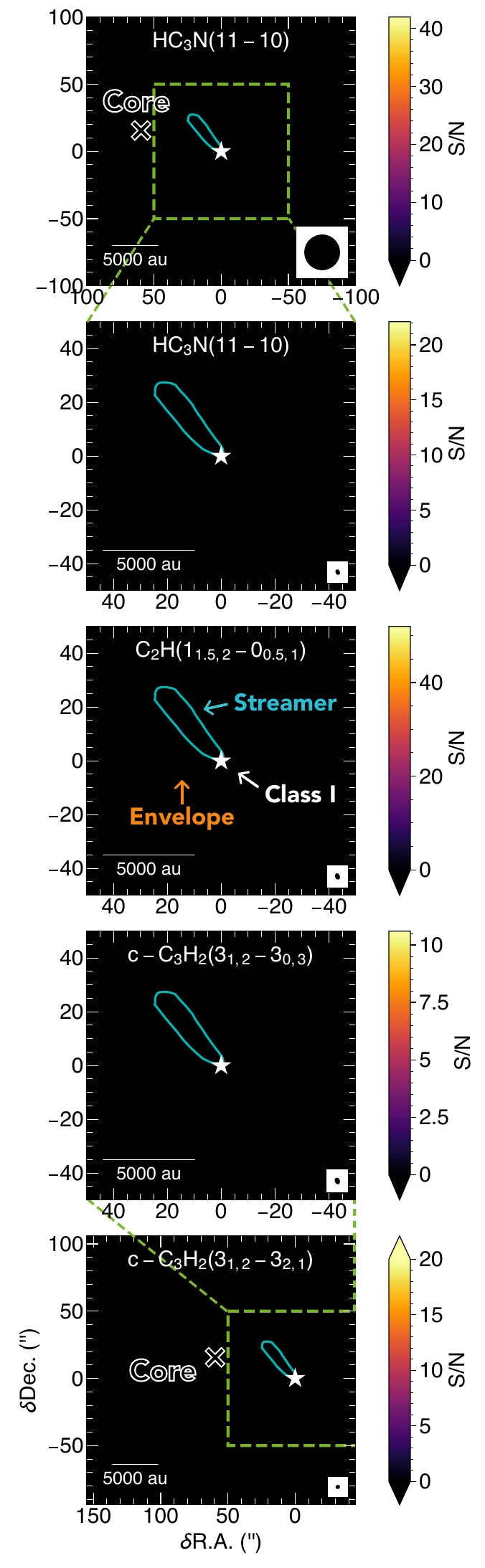}
    \caption{Gallery of S/N maps from few of the observed lines used in this work. The line name is indicated on the top of each panel while their beam is shown in the lower right corner. The star indicates the position of L1489~IRS. The cyan contour corresponds to the targeted emission for the analysis (see Sect.~\ref{sec:data-preprocessing}). The black cross indicates the position of the peak intensity of the core \citep{wu2019}. \textbf{Top to bottom:} IRAM-30m observations of \ce{HC3N} (first panel), single-field combined observations (IRAM-30m + NOEMA) of \ce{HC3N} (second), \ce{C2H} (third), and c-\ce{C3H2} (fourth), and mosaic combined observations of c-\ce{C3H2} (fifth).}
    \label{fig:snr-maps-simplified}
\end{figure}

The line properties used for the present study are listed in Table~\ref{table:lines}, while their integrated intensity maps are shown in Fig.~\ref{fig:mom0-maps}. We present in Fig.~\ref{fig:snr-maps-simplified} the S/N maps of few of our observed lines (see Fig.~\ref{fig:snr-maps} for all transitions), produced with the \texttt{GILDAS} software \texttt{CUBE}, which enabled us to enhance the structures. The latter were computed as the ratio of signal maps to the corresponding noise maps, accounting for the spatial variations of the noise. This allowed us to compare structures in terms of their detectability rather than pure emission, as the noise was not uniformly distributed; in mosaics, the sensitivity is variable due to the nonuniform and nonsimultaneous coverage of the observations and in single-pointing fields, the noise increases toward the edges because of the primary beam attenuation away from the pointing center. For Figs.~\ref{fig:snr-maps-simplified}, \ref{fig:mom0-maps}, and \ref{fig:snr-maps}, the maps were produced by integrating line intensity over the streamer velocity range, namely,~between 6.4 and 7.4~km~s$^{-1}$ derived in \citetalias{tanious2024} \citep[blue-shifted emission relatively to disk's LSR velocity of $\sim7.4$~km~s$^{-1}$,][]{yamato2023}. This results in emission localized on the eastern side of maps only. Small red-shifted \ce{C2H} and \ce{HC3N} emission are detected on the western side (see \citetalias{tanious2024}) but they are not relevant to this study and, thus, they are not presented here.

The IRAM-30m \ce{HC3N} emission (top panel of Fig.~\ref{fig:snr-maps-simplified}) probes a bright extended structure corresponding to the prestellar core L1489, which was previously observed in other carbon chains molecules \citep{wu2019}. The protostar L1489 IRS seems to lie at the edge of the core, but the connection between the two is not clear due to the poor spatial resolution. However, the c-\ce{C3H2} emission, observed with our mosaic observations, probes this connection through the streamer (bottom panel of Fig.~\ref{fig:snr-maps-simplified}). The position of the peak intensity in the core is similar in terms of \ce{HC3N} and mosaic c-\ce{C3H2} emission and it corresponds to the position of the peak intensity of other carbon species \citep[black cross on Fig.~\ref{fig:snr-maps-simplified},][]{wu2019}. Single-field observations of \ce{HC3N}, \ce{C2H}, and c-\ce{C3H2} (see the three middle panels of Fig.~\ref{fig:snr-maps-simplified}) probe a conical flow tracing infalling envelope oriented toward the core, whose northern part was identified as a streamer in \citetalias{tanious2024}.

%%%%%%%%%%%%%%%%%%%%%%%%%%%%%%%%%%%%%%%%%%%%%%%%%%%%%%%%%%%

\section{Estimation of column densities and abundances}
\label{sec:methods}

\subsection{Goals}
\label{sec:goals}
Here, we aim to estimate the streamer properties, namely, its mass and infall rate, along with its impact on the \{star+disk\} mass. This requires us to determine the total \ce{H2} column density map over the streamer, since summing the \ce{H2} column density on each pixel multiplied by the pixel size gives the number of \ce{H2} molecules in the streamer. It can then be converted to its mass assuming the molecular weight of the gas. Accordingly, we followed Eq.~(1) of \cite{valdivia-mena2022}, expressed as
\begin{equation}
    M_\text{streamer} = M_\text{gas} = \mu_\text{mol} m_\text{H} \, d^2 \sum_\text{pixels} \delta x \,\delta y \, N^{(x,y)}_{\text{H}_2}
    \label{eq:streamer-mass}
,\end{equation}
where $\mu_\text{mol}=2.8$ is the molecular weight of the gas considering contribution from hydrogen, helium, and metals \citep{kauffmann2008}, $m_\text{H}$ is the mass of an hydrogen atom, $d$ is the distance to the protostar in cm \citep[taken as 146~pc converted into cm,][]{roccatagliata2020}, $\delta x \,\delta y$ is the pixel size in radians, and $N^{(x,y)}_{\text{H}_2}$ is the \ce{H2} column density of the streamer in cm~$^{-2}$ in the pixel $(x,y)$. Since \ce{H2} is not observable at millimeter wavelengths (being a homonuclear molecule with no permanent dipolar moment), a proxy to access $N_{\text{H}_2}$ is to convert the total column density of a given molecule observable at millimeter waves using the molecular abundance with respect to \ce{H2}, using
\begin{equation}
    \NHtwo = \Ntot^\text{mol}/X
    \label{eq:Ntot_to_NH2}
,\end{equation}
where $\Ntot^\text{mol}$ is the total column density of a given molecule \texttt{mol}, and $X$ is the molecular abundance of \texttt{mol} wrt \ce{H2}.

We describe in Sects.~\ref{sec:hfs} and \ref{sec:RD} the two methods at local thermodynamic equilibrium (LTE) we used to derive the total column density maps of molecular species. In Sect.~\ref{sec:mcmc-radex}, we describe a third, non-LTE method to benchmark molecular column densities and obtained molecular abundances with respect to \ce{H2} from our observations. This gave us access to the streamer mass via Eqs.~\eqref{eq:streamer-mass} and \eqref{eq:Ntot_to_NH2}.

\subsection{Data pre-processing}
\label{sec:data-preprocessing}

The methods described below in this section are based on a pixel-by-pixel approach. Cubes used in the same analysis should have the same spatial properties to ensure a consistent pixel area over the beam area ratio,  getting consistent fluxes for all line cubes of the same molecule. All steps described in this subsection were performed using \texttt{CUBE}.
For a given molecule, we  smoothed each line cube to match the coarsest line cube's beam (i.e., $2\farcs12\times1\farcs47$ at $-157.6\degr$ for \ce{C2H} and $35\farcs63\times35\farcs63$ at $0\degr$ for \ce{HC3N}, cf. Table~\ref{table:lines}). In case of ambiguity, namely, the cube with the largest beam minor axis is not the one with the largest beam major axis (e.g., for c-\ce{C3H2}), we used a circular beam using the largest beam major axis (i.e., $2\farcs78\times2\farcs78$ at $0\degr$ for c-\ce{C3H2}). Then, for each molecule, we re-gridded its corresponding line cubes using the largest pixel size among the line cubes. 

Since we were focused on the streamer, we extracted after the steps above only its emission in each line cube, using the polygon shown in cyan in Fig.~\ref{fig:snr-maps-simplified} (from \citetalias{tanious2024}), which encompasses all the streamer emission. Eventually, we produced the integrated intensity maps using the streamer velocity range between 6.4 and 7.4~km~s$^{-1}$ derived in \citetalias{tanious2024}. Uncertainties on the integrated intensities were calculated as the quadrature of the zeroth-moment map's rms and the flux calibration uncertainty, taken to be 10\%.

\subsection{HFS fitting}
\label{sec:hfs}
Assuming the hyperfine structure (HFS) of an emission line is at LTE, we can access simultaneously to the opacity and the excitation temperature of the line and, thus, to the total column density of the molecule. Hyperfine resolved lines are thus a powerful tool to estimate column densities. 

We performed the fitting in \texttt{CUBE} using the \texttt{GILDAS} HFS method. \texttt{CUBE} HFS is based on a pixel-by-pixel approach, meaning that no spatial information of the neighbor pixels was used to fit parameter values of the current pixel. The fit uses a parameter file describing the HFS of the line, namely, the velocity (or frequency) offset relative to the main component, and the relative normalized intensity of each component. The latter for the $k$\textsuperscript{th}-component is computed as $\Aij^k g_\text{up}^k/(\sum \Aij g_\text{up})$, where $\Aij^k$ and $g^k_\text{up}$ are the Einstein coefficient and upper level degeneracy of the $k$\textsuperscript{th}-transition, respectively.

Four parameters are minimized in order to reproduce the observations: the systemic velocity of the main component, $v_\text{LSR}$; the full width at half maximum of the components, FWHM (assumed to be identical for each component); the total opacity of the group, $\tau$; and the product of the observed peak line intensity by the total opacity of the group, $T_\text{obs} \times \tau$ (which serves as an intermediary parameter to obtain the excitation temperature). Details on the implementation of HFS in \texttt{CUBE} can be found in Appendix B.1. of \citet{beslic2024}. After providing a global first guess to the algorithm (same value in every pixel), a few iterations were necessary to reach good convergence. In practice, the number of iterations was set to five, which proved sufficient for good convergence. 
As the output, we obtained four maps (cf. Fig~\ref{fig:c2h-hfs-results}), each corresponding to one parameter (along with four more corresponding to their error) of the fit.\\

Those maps enable to access the excitation temperature $\Tex$ and total column density $\Ntot$ maps. Indeed, by inverting the radiative transfer equation:
\begin{equation}
    T_\text{obs} = f[J_\nu(\Tex) - J_\nu(\Tbg)](1- e^{-\tau})
    \label{eq:TR_Jnu}
,\end{equation}
where $f$ is the beam filling factor, $\Tbg$ is the background temperature, and $J_\nu(T)$ is the Rayleigh-Jeans equivalent temperature defined as
\begin{equation}
    J_\nu(T) = \dfrac{h\nu/\kB}{e^{h\nu/\kB T}-1}
    \label{eq:Jnu_def}
,\end{equation}
where $h$ is the Planck constant, and $\kB$ is the Boltzmann constant, we can obtain the $\Tex$ map using
\begin{equation}
    \Tex = \dfrac{h\nu}{\kB}\left[ \ln\left(1+\dfrac{h\nu/\kB}{\frac{T_\text{obs}}{f(1-\exp(-\tau))} + J_\nu(\Tbg)}\right) \right]^{-1}
    \label{eq:Tex}
.\end{equation}
In the case of too optically thin emission, namely, $\tau<0.3$, the fit cannot converge properly as $\tau$ is simplified and disappears from equations (and, thus, it cannot be constrained; cf. \texttt{CLASS} documentation). 
Therefore, the $\Tex$ values for $\tau<0.3$ were masked. 
Finally, we used the general formula to compute the column density, $\Ntot$, map,  derived from Eq. (84) of \citet{mangum2015}:
\begin{equation}
    \begin{split}
        \Ntot(\Tex) & = \dfrac{Q(\Tex)}{g_\text{up}}  \dfrac{8\pi\nu^3}{\Aij c^3} \dfrac{\exp(E_\text{up}/\Tex)}{\exp(h\nu/\kB\Tex) - 1} \\
        & \times \int - \ln \left(1-\dfrac{T_\text{obs}(v)}{f[J_\nu(\Tex) - J_\nu(\Tbg)]}\right) \mathrm{d} v
    \end{split}
    \label{eq:Ntot_general}
,\end{equation}
where $c$ is the light velocity, $Q$ is the partition function of the molecule, $g_\text{up}$ and $E_\text{up}$ are the degeneracy and the energy of the upper level considered, $\nu$ and $A_{ij}$ are the frequency and the Einstein coefficient for spontaneous emission of the line, and $T_\text{obs}(v)$ is the line temperature profile.

\subsection{Rotational diagram}
\label{sec:RD}
Another common method to derive excitation temperature and column densities is the rotational or population diagram analysis \citep{goldsmith1999}. This method is based on the following LTE population equation,
\begin{equation}
    \ln\left(\dfrac{N_\text{up}}{g_\text{up}}\right) = -\dfrac{E_\text{up}}{\kB\Tex} + \ln(\Ntot) - \ln(Q(\Tex))
    \label{eq:rotational_diagram}
,\end{equation}
where $N_\text{up}$, $g_\text{up}$, and $E_\text{up}$ are the column density, the degeneracy, and the energy of the upper level considered, $\Tex$ is the excitation temperature, and $\Ntot$ and $Q$ are the total column density and the partition function of the molecule, respectively.

In the optically thin case and in the Rayleigh-Jeans regime, assuming also no significant emission from the background, we can use the usual simple relationship, 
\begin{equation}
    N^\text{thin}_\text{up} = \dfrac{8\pi\kB\nu^2}{\Aij fhc^3} \, M_0^{u \to l}
    \label{eq:Nu_optically_thin_RJ}
,\end{equation}
where $M_0^{u \to l}$ is the integrated intensity of the upper $\to$ lower line considered. 
This relation is in fact valid even outside the Rayleigh-Jeans regime (which is the case for millimeter observations), as we retrieved similar total column densities and excitation temperatures using a rotational diagram or a radiative transfer model codes (e.g., \texttt{RADEX}, see Table~\ref{table:summary-TR-results} and Sect.~\ref{sec:results}).

In the case of moderately optically thick lines, the column density estimated with the optically thin regime $N^\text{thin}_\text{up}$ can be corrected using the optical depth correction factor $C_{\tau_{ul}}$ for each transition $u \to l$, to get the actual column density of the upper state $N_\text{up}$, through $N_\text{up} = N^\text{thin}_\text{up} \times C_{\tau_{ul}}$. This factor  \citep{goldsmith1999} is defined as
\begin{equation}
    C_{\tau_{ul}} = \tau_{ul}/(1 - e^{-\tau_{ul}})
    \label{eq:Ctau}
.\end{equation}
The opacity value can be accessed through an inversion of Eq.~\eqref{eq:TR_Jnu}, which gives
\begin{equation}
    \tau_{ul} = - \ln\left(1-\dfrac{T^{u\to l}_\text{obs}}{f[J_\nu(\Tex) - J_\nu(\Tbg)]}\right)
    \label{eq:tau_ln}
,\end{equation}
which assumes that the excitation temperature, $\Tex$, is known. Therefore, we iterated multiple times Eq.~\eqref{eq:rotational_diagram} using on the first iteration $C_{\tau_{ul}}=1$ (i.e., optically thin regime) for all transitions. We then computed $C_{\tau_{ul}}$ in the following iterations, using Eqs.~\eqref{eq:Ctau} and \ref{eq:tau_ln}, along with $\Tex$ derived from Eq.~\eqref{eq:rotational_diagram}. Iterations on $\tau_{ul}$ were performed until a convergence better than 5\% was reached.

Multiple tests using only emission above a certain S/N showed that using emission with S/N~$\geq 3$ leads to the best results (i.e., obtaining a continuous and consistent map, while keeping enough pixels). For robustness, we retained  only those pixels with three lines or more.
Using Eq.~\eqref{eq:Nu_optically_thin_RJ} allows us to directly infer the upper column densities, $N_\text{up}$, from the observed integrated intensities and access to the excitation temperature (assumed to be the same for all levels), as well as the total column density of the molecule using Eq.~\eqref{eq:rotational_diagram}.

\subsection{Abundances derivation with MCMC \texttt{RADEX}}
\label{sec:mcmc-radex}

As a third method, we used the non-LTE \texttt{RADEX} code \citep{vandertak2007} to reproduce the observed integrated line intensities. This code takes as input an escape probability method, a molecular collisional data file, a kinetic temperature, $\Tkin$, a background temperature, $\Tbg$, a molecular total column density, $\Ntot$, a line width (assumed to be the same for all transitions), and a number of collisional partners along with their respective densities. In the present work, only collisions with para-H$_2$ were considered since it is the dominant collider in cold and dense gas \citep[see e.g.,][]{troscompt2009}. 
Thus, in contrast to the two previous methods, this non-LTE approach provides access to the density of \ce{H2}, which is a critical parameter for determining the streamer mass.

To best explore the parameter space, we used the Markov chain Monte Carlo  (MCMC) approach implemented in the python package \texttt{emcee} \citep{emcee}, which we coupled to \texttt{RADEX}, following methodologies commonly used in the literature \citep[e.g.,][]{yang2017,Hily2018,loomis2021,xue2024,sil2025}. The MCMC algorithm produces a vector of parameters for which the likelihood of representing the ``true'' physical conditions is estimated from the ability of the model to reproduce the observed lines intensities.  
To couple it with \texttt{RADEX}, we developed a python wrapper consisting in translating the vector as an input file executable by \texttt{RADEX}. The output from \texttt{RADEX} is parsed by our code, which estimates the likelihood for that set of results. This process was repeated iteratively to explore the parameter space. 

We used the average of each integrated line intensity across the entire streamer as representative inputs for the model. Indeed, tests consisting in running the code in two parts of the streamer (i.e., averaged emission of the upper part vs. the lower part) showed no significant differences on the results within the uncertainties of the derived parameters. Hence, we used the globally averaged integrated emission from the whole streamer as input for the MCMC \texttt{RADEX} analysis.

To reproduce observed integrated intensities, we used the uniform sphere method for the escape probability\footnote{The use of the slab prescription showed no significant differences on the results.}, varied the kinetic temperature, $\Tkin$, the total column density, $\Ntot$, and the para-\ce{H2} density, \npH, and fixed the line width at 0.5~km~s$^{-1}$ (derived from observed spectra) as well as the background temperature, $\Tbg$, to 2.73~K. We used 32 walkers on 10~000 steps to ensure the convergence. They were  initialized and set to explore the ranges reported in Table~\ref{table:parameters-mcmc}. The initial values for $\Tex$ and $\Ntot$ were taken from results of the HFS fitting or the rotational diagram. The range of the para-\ce{H2} density was defined from typical values found in molecular clouds ($10^3-10^6$~cm$^{-3}$), which were consistent with the expected \ce{H2} density derived from the computation of the critical densities of the observed lines (see Appendix~\ref{sec:critical-densities}). The collisional data were taken from \cite{Pirlot2023} for C$_2$H, \cite{BenKhalifa2019} for ortho-c-C$_3$H$_2$ and \cite{Hily2018} for HC$_3$N, as compiled in the EMAA database\footnote{\url{https://emaa.osug.fr}} \citep{emaa}.  

This method also allowed us to estimate the kinetic temperature, which could be compared to the excitation temperature derived either from the HFS fit or the rotational diagram methods described above. Therefore, it  helps to assess whether the LTE assumption holds. It also provides an estimate of the total column density of molecule \texttt{mol}, $\Ntot^\text{mol}$, enabling a comparison across the different methods to check for consistency. Additionally, it yields the para-\ce{H2} density \npH, which is of main interest to estimate the streamer mass. Assuming the depth of the streamer corresponds to its width, $w$ (taken as 8\arcsec, converted in cm), and assuming $n_{\text{H}_2} = n_{\text{p-H}_2}$ (in cm~$^{-3}$, as \ce{H2} is mainly in para form in the cold interstellar medium, see above), we can then derive the molecular abundance with respect to \ce{H2} of molecule \texttt{mol},
\begin{equation}
    X = \left[ \dfrac{n_\text{mol}}{n_{\text{H}_2}} \right]_\text{streamer} = \dfrac{\Ntot^\text{mol}}{n_{\text{H}_2}\,w}
    \label{eq:abundance}
.\end{equation}

As we averaged integrated emission over the streamer, we obtained a mean abundance over the streamer for each molecular species. The latter was then used in Eq.~\eqref{eq:Ntot_to_NH2} for each pixel of the streamer column density maps, derived with the HFS fit (Sect.~\ref{sec:hfs}) or the rotational diagrams (Sect.~\ref{sec:RD}), to get a \ce{H2} streamer column density map. The latter was then converted to a mass following Eq.~\eqref{eq:streamer-mass}, as detailed in Sect.~\ref{sec:goals}.

\begin{table*}[!ht]
    \caption{Parameters used for the MCMC \texttt{RADEX} exploration.}    
    \label{table:parameters-mcmc}      
    \centering    
    \renewcommand{\arraystretch}{1.5} 
    \begin{tabular}{c c c c}        
    \hline\hline                 
        Parameter & \ce{C2H} & o-c-\ce{C3H2} & \ce{HC3N} \\
    \hline    
        $\Tkin$ (K)  & 7.3, $[5 - 10]$ & 7.8, $[5 - 20]$ & 6.1, $[5 - 11]$ \\
        
        \npH (cm$^{-3}$) & $10^5$, $[10^3 - 10^6]$ & $10^5$, $[10^3 - 10^6]$ & $10^5$, $[10^3 - 10^6]$  \\
        
        $\Ntot$ (cm$^{-2}$) & \num{1.83e14}, $[10^{13} - 10^{15}]$ & \num{3.44e12}, $[10^{11} - 10^{13}]$  & \num{3.25e13}, $[10^{12} - 10^{14}]$\\
        
        $\Tbg$ (K)   & 2.73 & 2.73 & 2.73 \\
        FWHM (km~s$^{-1}$) & 0.5 & 0.5 & 0.5 \\
    \hline                                   
    \end{tabular}
    \tablefoot{If a range is provided for a parameter, the MCMC algorithm explores the range starting with the initial value in front of the range. If a single value is indicated, the parameter is fixed to this value.}
\end{table*}

%%%%%%%%%%%%%%%%%%%%%%%%%%%%%%%%%%%%%%%%%%%%%%%%%%%%%%%%%%%

\section{Results}
\label{sec:results}

\subsection{\ce{C2H}}
\label{sec:results-c2h}

\begin{figure}
    \centering
    \includegraphics[width=\linewidth]{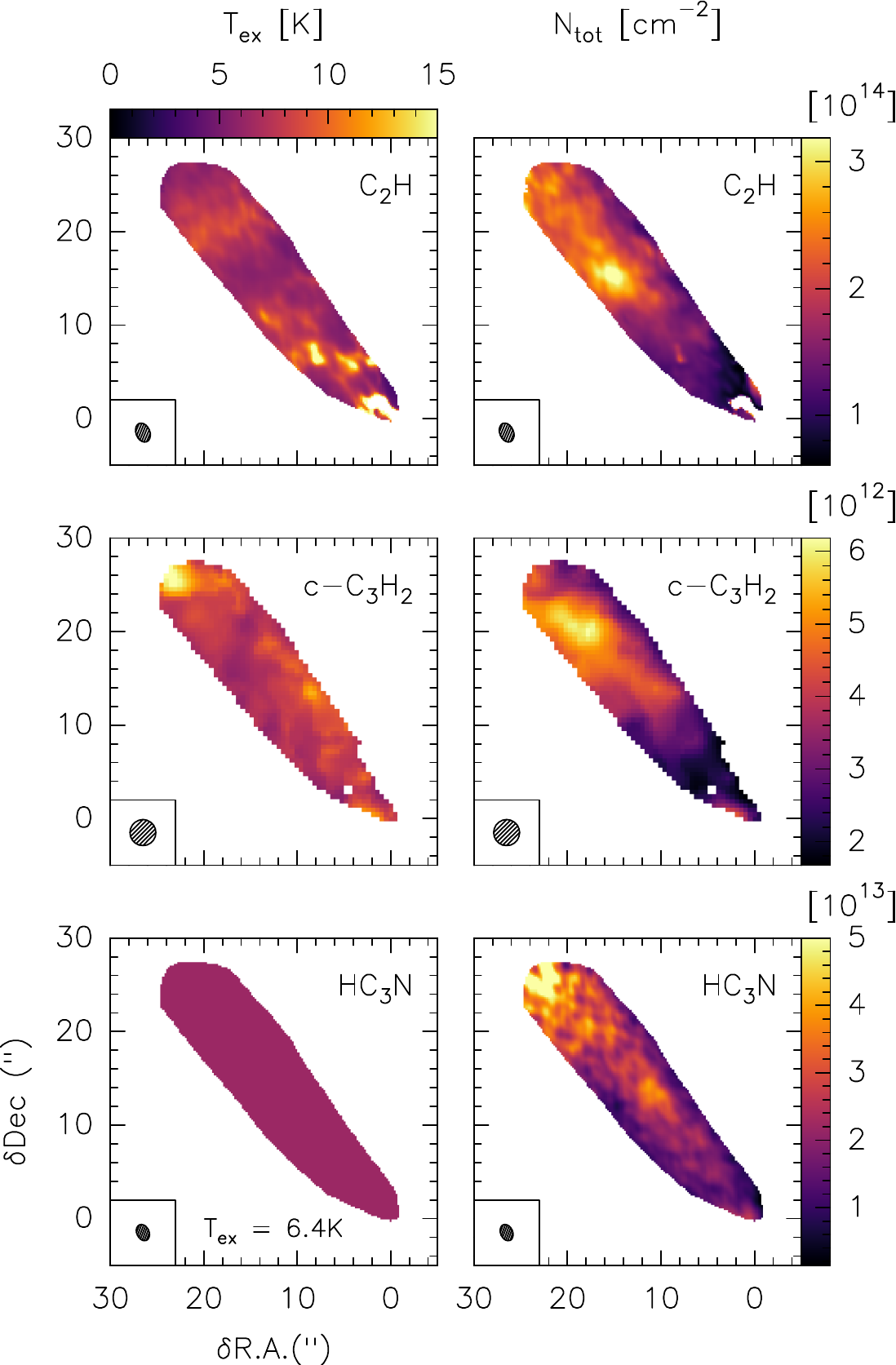}
    \caption{Excitation temperature (left column) and column density (right column) maps for \ce{C2H} (top), ortho-c-\ce{C3H2} (middle), and \ce{HC3N} (bottom), derived with methods described in the text. The beam displayed on the bottom left corner corresponds for \ce{C2H} and c-\ce{C3H2} to the smoothed beam used in the analysis (see Sect.~\ref{sec:data-preprocessing}), and to the original beam from NOEMA observations for \ce{HC3N} (see Sect.~\ref{sec:results-hc3n}).
    }
    \label{fig:Tex_Ntot}
\end{figure}

We took advantage of the \ce{C2H}~$(J=1\to0)$ line whose HFS is composed of six different components that are easily resolved thanks to their large velocity offsets. We applied the HFS fitting method described in Sect.~\ref{sec:hfs}, using an initial guess for $T_\text{obs}=4$~K, $v_\text{LSR}=6.8$~km~s$^{-1}$, FWHM$=0.5$~km~s$^{-1}$, and $\tau=5$. The maps of fitted parameters are shown in Fig.~\ref{fig:c2h-hfs-results}. The fit reproduces  the observed intensities well, with the maximum normalized $\chi^2=0.19$. From Eqs.~\eqref{eq:Tex} and \eqref{eq:Ntot_general}, using the best fitted component, $\Tbg=2.73$~K, and $f=1$ (noting that the streamer was resolved with NOEMA observations), then interpolating the partition function of \ce{C2H} from the CDMS at low temperatures, we accessed the excitation temperature and total column density maps of \ce{C2H} (shown in the top row of Fig.~\ref{fig:Tex_Ntot}). The excitation temperature is rather flat over the streamer, with 68\% of values ranging between 5.8~K (16th percentile) and 8.6~K (84th percentile), and a mean value at 7.3~K. The same trend is seen for the column density, with a mean over the streamer of $\Ntot=(1.83\pm0.56) \, \times 10^{14}$~cm$^{-2}$.

The \ce{C2H} abundance was determined using the MCMC \texttt{RADEX} method described in Sect.~\ref{sec:mcmc-radex}, applied to the integrated intensities extracted from observed line cubes averaged over the entire streamer emission. The initial kinetic temperature and column density were set to the mean excitation temperature and mean column density derived with the HFS fit. 
For the kinetic temperature, the range was reduced from $[5-20]$~K to $[5-10]$~K after few tests showing no good convergence of the model in the range of $[10-20]$~K. The corner plot after removing 1500 steps of burn-in is shown in Fig.~\ref{fig:corner-plot-c2h}, while best modeled (i.e., the 50th-percentile model) integrated intensities and excitation temperatures are given in Table~\ref{table:c2h-obs_vs_RADEX}. 
The best model gives $\Tkin=7.3^{\,+1.3}_{\,-0.7}$~K and $\Ntot=(2.18^{\,+0.31}_{\,-0.27})\times 10^{14}$~cm$^{-2}$. Meanwhile, the mean value of the \npH density is poorly constrained, but its lower bound appears to be well determined. Thus, we derived a lower limit of \npH$\geq \num{1.31e5}$~cm$^{-3}$ corresponding to the highest probability value.
Using Eq.~\eqref{eq:abundance}, we ended with an upper limit for the \ce{C2H} abundance in the streamer of $\leq \num{1.09e-7}$, resulting in a lower limit for the \ce{H2} column density of $\NHtwo \geq 1.17 \times 10^{21}$~cm$^{-2}$ (from Eq.~\eqref{eq:Ntot_to_NH2}). 

\begin{table*}
    \caption{Summary of radiative transfer model results on the streamer.}   
    \small    
    \label{table:summary-TR-results}      
    \centering                          
    \renewcommand{\arraystretch}{1.8} 
    
    \begin{tabular}{l c c c c c l}
    \hline
         Species   &   $\Ntot$   &   $\Tex$   &   $\Tkin$   &   \npH   &   Abundance $X$   &   Method   \\
                   &(cm~$^{-2}$) &     (K)    &      (K)    & ($10^5$~cm~$^{-3}$) &               &            \\
    \hline
    \hline
         \ce{C2H}  &  $(1.83\pm0.56) \, \times 10^{14}$  &  $7.3^{\,+1.3}_{\,-1.5}$ &  ---  &  ---  &  ---  &  HFS fitting \\
         
                   &  $(2.18^{\,+0.31}_{\,-0.27})\,\times 10^{14}$  &  6.6\tablefootmark{(a)} &  $7.3^{\,+1.3}_{\,-0.7}$  &  $\geq1.31$   & $\leq \num{1.09e-7}$   &  MCMC \texttt{RADEX}  \\
                   
    \hline
    o-c-\ce{C3H2}  &  $(3.44^{\,+1.41}_{\,-0.95})\,\times 10^{12}$  &  $7.8^{\,+1.4}_{\,-1.0}$ &  ---  &  ---  &  ---  &  Rotational diagram \\
    
                   &  $(3.28^{\,+0.33}_{\,-0.28})\,\times 10^{12}$  &  10.1\tablefootmark{(a)} &  $12.4^{\,+4.0}_{\,-2.3}$ &  $\geq 3.77$   & $\leq \num{5.48e-10}$   &  MCMC \texttt{RADEX} \\
                   
    \hline
        \ce{HC3N}  &  $(3.25\pm1.09) \, \times 10^{13}$  &  $6.1\pm0.5$ &  ---  &  ---  &  ---  &  Rotational diagram on 30m lines \\
        
                   &  $(3.05^{\,+1.89}_{\,-1.04})\,\times 10^{13}$  &  6.3\tablefootmark{(b)} &  $6.8^{\,+1.0}_{\,-0.6}$  & $\geq 1.99$   & $\leq \num{1.42e-8}$    &  MCMC \texttt{RADEX} on 30m lines \\
                   
                   &  $(2.39\pm0.99)\,\times 10^{13}$  &  6.3 &  ---  &  ---  &  ---  &  Using Eq.~\eqref{eq:Ntot_general} and $\Tex=6.3$~K for NOEMA's line \\
    \hline
    \end{tabular}
    \tablefoot{
        \tablefoottext{a}{Averaged on all levels from the best \texttt{RADEX} model.}
        \tablefoottext{b}{ $J=11-10$ level from the best \texttt{RADEX} model.}
        }
\end{table*}

\subsection{Ortho-c-\ce{C3H2}}
\label{sec:results-ocC3H2}

We applied the rotational diagram method described in Sect.~\ref{sec:RD}, pixel by pixel, on four ortho lines of c-\ce{C3H2} to produce the excitation temperature and column density maps shown in the middle row of Fig.~\ref{fig:Tex_Ntot}. Across the streamer, the excitation temperature or column density exhibits no substantial variation with $\Tex=7.8^{\,+1.4}_{\,-1.0}$~K and $\Ntot=(3.44^{\,+1.41}_{\,-0.95})\,\times 10^{12}$~cm$^{-2}$.

Using the MCMC \texttt{RADEX} on the observed averaged integrated intensities with initial values of $\Tkin$ and $\Ntot$ set by the rotational diagram, we obtained the corner plot shown in Fig.~\ref{fig:corner-plot-ocC3H2} after removing 1500 burn-in steps. Thus, we obtained the intensities in Table~\ref{table:cc3h2-obs_vs_RADEX_allres}. Similarly to \ce{C2H}, we inferred the kinetic temperature $\Tkin=12.4^{\,+4.0}_{\,-2.3}$~K, another estimation of the column density $\Ntot=(3.28^{\,+0.33}_{\,-0.28})\,\times 10^{12}$~cm$^{-2}$, a lower limit on the para-\ce{H2} density \npH$\geq \num{3.77e5}$~cm$^{-3}$, and eventually an upper limit on the abundance of ortho-c-\ce{C3H2} of $\leq \num{5.48e-10}$. This leads to an estimation of the lower limit on the column density of \ce{H2} over the streamer of $\NHtwo \geq \num{4.55e21}$~cm$^{-2}$.

\subsection{\ce{HC3N}}
\label{sec:results-hc3n}

\begin{figure}
    \centering
    \includegraphics[width=\linewidth]{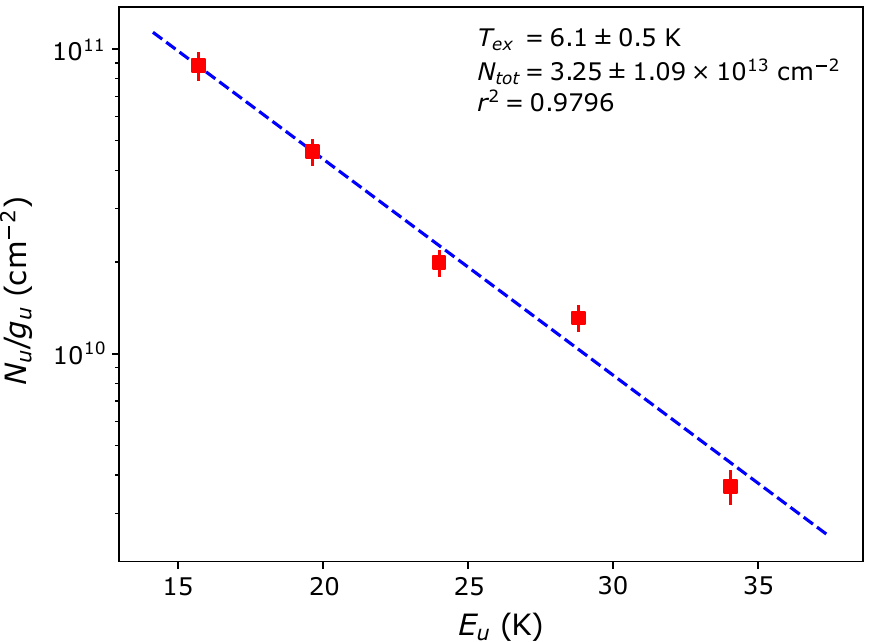}
    \caption{Rotational diagram of \ce{HC3N} on observed IRAM-30m lines toward the streamer. Red points correspond to the observations while the blue dashed line correspond to the best linear fitting.
    }
    \label{fig:hc3n-rd}
\end{figure}

Unlike \ce{C2H} or c-\ce{C3H2}, our current dataset contains only a single line of \ce{HC3N} resolved with NOEMA, preventing us from using radiative transfer methods. Nevertheless, we were still able to observe five lines within our IRAM-30m dataset. We took advantage of those observations to infer a mean excitation temperature and column density of \ce{HC3N} with the rotational diagram shown in Fig.~\ref{fig:hc3n-rd}, using the emission of the single pixel over the streamer. As this structure is diluted in one IRAM-30m beam, but resolved with NOEMA, we corrected the emission by the beam filling factor $f$ computed as the ratio of the streamer's area by the beam's area, leading to $f=0.253$. The estimated mean excitation temperature is $\Tex=6.1\pm0.5$~K, while the mean column density is $\Ntot=(3.25\pm1.09) \, \times 10^{13}$~cm$^{-2}$.

Those results were used as the initial values for the kinetic temperature and column density in the MCMC \texttt{RADEX} model to reproduce IRAM-30m line integrated intensities corrected by the beam filling factor. Similarly to \ce{C2H}, the kinetic temperature range was narrowed from $[5-20]$~K to $[5-11]$~K, following preliminary tests showing poor convergence at higher temperatures. The outputs are presented in Fig.~\ref{fig:corner-plot-hc3n} and Table~\ref{table:hc3n-obs_vs_RADEX_allres}. The kinetic temperature is estimated to be $\Tkin=6.8^{\,+1.0}_{\,-0.6}$~K, the column density $\Ntot=(3.05^{\,+1.89}_{\,-1.04})\,\times 10^{13}$~cm$^{-2}$, and the lower limit on the para-\ce{H2} density \npH$\geq \num{1.99e5}$~cm$^{-3}$. The upper limit on the \ce{HC3N} abundance is therefore estimated to be $\leq \num{1.42e-8}$. 

The best \texttt{RADEX} model also provides the excitation temperature for each transition. Thus, we applied Eq.~\eqref{eq:Ntot_general}, with the computed $\Tex$ for the $J=11\to10$ level, on the NOEMA resolved line to obtain a resolved column density map of \ce{HC3N}, which is presented in the bottom row of Fig.~\ref{fig:Tex_Ntot}. The mean column density over the streamer is $\Ntot=(2.39\pm0.99)\,\times 10^{13}$~cm$^{-2}$. Given the derived abundance, the lower limit on the \ce{H2} column density is estimated as $\NHtwo \geq \num{9.90e20}$~cm$^{-2}$.

\subsection{Streamer properties}
\label{sec:discussion-streamer-properties}

\begin{table}
    \caption{
    Summary of \ce{H2} column densities, streamer mass, infall rate, and streamer mass relative to the disk's mass, using the \texttt{TIPSY} modeled infalling time of 24.1~kyr. 
    }   
    
    \small
    
    \label{table:summary-mass-results}      
    \centering                          
    \renewcommand{\arraystretch}{1.4} 
    
    \begin{tabular}{c c c c}    
    \hline\hline                 
        Tracer & \ce{C2H} & o-c-\ce{C3H2}  &  \ce{HC3N} \\
                 
    \hline 
        \begin{tabular}{@{}c@{}}$N_{H_2}$ \\ ($10^{21}$~cm~$^{-2}$)\end{tabular}
         & $\geq 1.17$ & $\geq 4.55$ & $\geq 0.99$ \\
        
    \hline  
        \begin{tabular}{@{}c@{}}Streamer mass \\ ($10^{-3}$~$M_\odot$)\end{tabular}
         & $\geq 4.67$ & $\geq 18.3$ & $\geq 4.76$ \\
        
    \hline 
        \begin{tabular}{@{}c@{}}Infall rate \\ ($10^{-7}$~$M_\odot$/yr)\end{tabular}
         & $\geq 1.94$ & $\geq 7.57$ & $\geq 1.97$ \\
        
    \hline 
        \begin{tabular}{@{}c@{}}Relative mass\tablefootmark{(\dag)} \\ (\%) \end{tabular}
         & $\geq 65$ & $\geq 257$ & $\geq 67$ \\
    \hline 
    \end{tabular}

    \tablefoot{\\
        \tablefoottext{\dag}{with respect to the disk's mass $M_\text{disk}=\num{7.1e-3}\,M_\odot$ \citep{sai2020}.}
    }
\end{table}

All results described above are summarized in Table~\ref{table:summary-TR-results}. From the \ce{H2} column density and abundances derived for each molecular tracer, we can infer a lower limit on the streamer mass, estimated as $\geq (4.67-18.3)\, \times 10^{-3}$~$M_\odot$ using Eq.~\eqref{eq:streamer-mass}, with the results given in Table~\ref{table:summary-mass-results}. The range corresponds to the three tracers (\ce{C2H}, ortho-c-\ce{C3H2,} and \ce{HC3N}), where ortho-c-\ce{C3H2} provides the higher upper limit, which we found to be the most reliable (see Sect.~\ref{sec:discussion-h2-density}). Assuming an infalling time, we can access its infalling rate. The streamer was modeled in \citetalias{tanious2024} using the \texttt{TIPSY} package, which enabled us to fit position-position-velocity cubes from observed line emission with infalling models \citep[free-fall motion due to gravity of the central object within a rotating cloud,][]{tipsy2024}. 
It allowed us to derive streamer properties based on the best fit model, such as its trajectory, its velocity profile, or its infalling time, as well as errors on each of these parameters according to the fit quality.
From the best fit model of \citetalias{tanious2024}, we estimated an infalling time of $24.1\pm1.6$~kyr.

We derived a lower limit of the infalling rate of the streamer onto the protoplanetary disk of $\geq (1.94-7.57)\, \times 10^{-7}$~$M_\odot$~yr$^{-1}$. Finally, assuming that the system accretes all the mass contained in the streamer and using a disk mass of $M_\text{disk}=\num{7.1e-3}\,M_\odot$ \citep{sai2020}, the relative mass provided by the streamer to the disk would be $\geq 65-257$\%. Those results are summarized in Table~\ref{table:summary-mass-results}. 

We note that compared to standard methods \citep[e.g., assuming canonical abundances to derive the streamer mass like in][]{valdivia-mena2022}, the MCMC \texttt{RADEX} has the advantage of being self-consistent (i.e., with no assumptions on the abundances). The abundances were derived strictly from observations (see Sect.~\ref{sec:mcmc-radex}), providing the least biased estimate of the streamer mass.

%%%%%%%%%%%%%%%%%%%%%%%%%%%%%%%%%%%%%%%%%%%%%%%%%%%%%%%%%%%

\section{Discussion}
\label{sec:discussion}

\subsection{\ce{H2} density}
\label{sec:discussion-h2-density}

While kinetic temperatures and column densities are relatively well constrained by the individual MCMC models, \ce{H2} densities are poorly constrained (see corner plots in Appendix~\ref{sec:mcmc-results-appendix}), leading to significant uncertainties in the derived abundances. This is likely due to near-LTE conditions ($\Tex\simeq\Tkin$) under which line emissions are poorly sensitive to the \ce{H2} density, as observed  for \ce{C2H} and \ce{HC3N} in the present work (see Table~\ref{table:summary-TR-results}). In contrast, the differences between the derived excitation temperatures for ortho-c-\ce{C3H2} lines (cf. Table~\ref{table:cc3h2-obs_vs_RADEX_allres}) indicate a clear departure from LTE, with deviations up to 6.1~K between the $2_{1,2}-1_{0,1}$ and $4_{1,4}-3_{0,3}$ lines. This suggests that the lower limit on the \ce{H2} density is better constrained by ortho-c-\ce{C3H2} compared to \ce{C2H} and \ce{HC3N} that are much closer to LTE. 
Nevertheless, this estimate could be significantly improved, either by providing additional (non-LTE) transitions to help the MCMC \texttt{RADEX} to converge or by using a molecular tracer with higher critical densities ($\gtrsim10^{6}$~cm$^{-3}$, e.g., HCN, see Table~\ref{table:critical-densities}) that would better trace the $\sim10^{5}$~cm$^{-3}$ density regime, resulting in more precise constraints on the streamer properties.

Now, despite different excitation conditions, all three species lead to very similar lower limits of \ce{H2} density; namely,~a \ce{H2} column density of $\NHtwo\gtrsim 10^{21}$~cm$^{-2}$, which is the typical magnitude found in molecular clouds \citep{sanhueza2012}. More specifically, for \object{Barnard 207}, the parental molecular cloud of \object{L1489~IRS}, \ce{H2} column densities of $10^{21}-10^{22}$~cm$^{-2}$ were found from the 500~$\mu$m \textit{Herschel}-SPIRE map \citep{togi2017}. Our values are also in good agreement with determinations toward the nearby \object{L1489} core, where the \ce{H2} density was found to be $n_{\text{H}_2}=(5.8^{\,+0.3}_{\,-0.4})\times 10^{5}$~cm$^{-3}$ \citep[using an elaborate non-LTE HFS fitting of three lines of \ce{N2H+},][]{daniel2007} and the \ce{H2} column density $\NHtwo=(1.02\pm0.07)\times10^{22}$~cm~$^{-2}$ \citep[using a SED fit on the dust emission peak,][]{wu2019}. Moreover, as detailed in Sect.~\ref{sec:discussion-ntot-x}, the retrieved abundances based on our estimate of the \ce{H2} column density are close to the literature values for the three species. For all these reasons, this estimate of the \ce{H2} density appears to be robust.

\subsection{Molecular column densities and abundances}
\label{sec:discussion-ntot-x}

The best MCMC \texttt{RADEX} model for \ce{C2H} gives HFS excitation temperatures that are very close to each other (see Table~\ref{table:c2h-obs_vs_RADEX}). Moreover, $\Tex\simeq\Tkin$ and the similarity between the $\Tex$ of each transition are an indication of near-LTE conditions. Those two conditions support the assumptions used in the HFS fitting method. The fit quality is indeed high, with a maximum normalized $\chi^2=0.19$, providing strong confidence in the derived \ce{C2H} column density, which is consistently retrieved using both HFS fitting and MCMC \texttt{RADEX} methods. Comparable values have been reported in dense cores \citep{sanhueza2012,giers2023}, protostellar regions \citep[e.g., in the Perseus molecular cloud,][]{higuchi2018} and even in protoplanetary disks \citep{bergner2019}. 

Similarly to \ce{C2H}, we find that \ce{HC3N} shows near-LTE conditions with close $\Tex$ values for each line, validating the rotational diagram method. The \ce{HC3N} column density we derived in the streamer is close to the one derived in the L1489 core \citep{wu2019}. 
However, our \ce{HC3N} column density estimation is two orders of magnitude higher than previous single-pointing measurements with IRAM-30m toward L1489 IRS \citep{bergner2017,law2018}, which may have underestimated the column density due to beam dilution of the thin structure, resulting in a blending between the disk and streamer emission.
A \ce{HC3N} column density of the same magnitude as our estimate has been measured in the Class 0 protostar L1527, also considered as the warm carbon chains chemistry (WCCC) reference \cite[$\Ntot(\ce{HC3N})=(2.52\pm0.05)\times10^{13}$~cm$^{-2}$,][]{araki2016}. Such WCCC sources are objects where unsaturated carbon chains are produced via gas-phase reactions, initiated by the sublimation of \ce{CH4} at temperatures around 25–30~K \citep{sakai2008}. This supports the idea that the L1489 core is a WCCC object candidate, as previously reported \citep{wu2019}. However, temperatures derived in the streamer are well below the temperature range of WCCC (see Table~\ref{table:summary-TR-results}), favoring the inheritance scenario of \ce{HC3N} from the core to the streamer, rather than a \ce{CH4}-triggered chemical reprocessing in the streamer. Further observations of other carbon chain molecules are needed to solidly confirm this.

In contrast, ortho-c-\ce{C3H2} shows non-LTE effects in some regions, which can be assessed with the $r^2$ (coefficient of determination) map 
indicating the quality of the linear regression fit of the rotational diagram in each pixel (see Fig.~\ref{fig:rd-ocC3H2-r2}). The $r^2$ coefficient is an indicator quantifying if the linear fit is able to reproduce the variance of observed data. Its value ranges from 0 to 1. The closer the value is to 1, the better the fit. Therefore, the reliability of the rotational diagram may be questioned and a specific analysis of these regions using a non-LTE method may be necessary. Nevertheless, the mean column density derived from the LTE rotational method is consistent with the non-LTE MCMC \texttt{RADEX} modeling, suggesting minor effects on results regardless of the prescription used (i.e., LTE or non-LTE).
Toward the L1489 core, the ortho-c-\ce{C3H2} column density was estimated to be $(5.5\pm0.1)\times 10^{12}$~cm$^{-2}$ using the $2_{1,2}\to1_{0,1}$ ortho-transition and assuming a single $\Tex=12.6\pm1$~K \citep{wu2019}, which is in good agreement with our values.

The ortho-para ratio of c-\ce{C3H2} observed in dense cores varies between 1 and 3 \citep{park2006}, resulting in a total column density of c-\ce{C3H2} between 2 and 1.33 times higher than the ortho column density. Therefore, using values from the MCMC \texttt{RADEX} code (cf. Table~\ref{table:summary-TR-results}), 
the \ce{C2H}/c-\ce{C3H2} ratio is $26.4-62.4$, which is much higher than the ratio of $\sim10$ that is found in dense cores \citep{gerin2011,cuadro2015,higuchi2018}. Overall, \ce{C2H} is known to be efficiently formed in UV irradiated regions, such as that of the Orion bar PDR's edge, where the \ce{C2H}/c-\ce{C3H2} ratio is $\sim32$ \citep{cuadro2015}; alternatively, it can be formed in the cavity walls of outflows illuminated by the central star \citep[e.g.,][]{oya2014}. 
The same photochemical mechanism may be at the origin of the \ce{C2H} enhancement in the streamer.
However, this scenario seems unlikely, as the kinetic temperatures for such mechanism lie around $10^2$~K \citep{cuadro2015}, which are much higher than the temperature derived here for \ce{C2H} (cf. Table~\ref{table:summary-TR-results}).

In addition, \ce{C2H} is also commonly observed in dense cores that are well shielded from UV illumination, where it traces dense parts \citep{higuchi2018}. This raises the possibility that the observed \ce{C2H} in the streamer (or at least part of it) could, in fact, be inherited from the parental dense core, which would be enhanced in \ce{C2H} or depleted in c-\ce{C3H2}, and not issued from photochemical mechanism.
The inheritance scenario is further supported by the similarity of the \ce{HC3N}/c-\ce{C3H2} ratio between the streamer ($1.9-8.5$, using MCMC \texttt{RADEX} values, see Table~\ref{table:summary-TR-results}) and the core \citep[$3.7-6.9$,][]{wu2019}. To assess whether inheritance could serve as a viable explanation for those ratios, a detailed chemical analysis of the L1489 core will be the scope of a forthcoming paper. 
Finally, the molecular abundances derived from \ce{H2} column densities (see Table~\ref{table:summary-TR-results}) are close to those found in similar objects, such as clumps \citep[e.g., $X_{\ce{C2H}}=\num{3.72e-8}$ and $X_{\ce{HC3N}}=\num{4.23e-10}$, see][]{sanhueza2012}, or in protostar envelopes; for instance, the Class I protostar \object{IRAS 04181+2655} also located in Taurus \citep[e.g., $X_{\ce{C2H}}=(1.94\pm6.9)\times10^{-8}$, $X_{\ce{c-C3H2}}=(2.8\pm2.0)\times10^{-10}$, and $X_{\ce{HC3N}}=(1.0\pm0.9)\times10^{-8}$; see][]{zhang2021}.

\subsection{Streamer's impact on the \{star+disk\} system}
\label{sec:discussion-streamer-impact}
We obtained a streamer mass of $\geq \num{4.67e-3}$~$M_\odot$ from \ce{C2H} and \ce{HC3N}, which corresponds to $\geq23-15$\% of the mass remaining in the envelope around the YSO estimated in the literature \citep[$\sim0.02-0.03\,M_\odot$,][]{motte2001,sheehan2017}. Using the lower limit and more reliable estimation based on ortho-c-\ce{C3H2} of $\geq \num{1.83e-2}$~$M_\odot$, this value reaches $\geq 92-61$\%. To better estimate what fraction of the envelope mass is contained in the streamer, a more precise estimate of the envelope mass is required. This could be achieved by applying the same methods to the envelope as those used to calculate streamer mass, but this requires a geometric model of the envelope. As the estimate of molecular abundances in a structure is based on Eq.~\eqref{eq:abundance}, the knowledge of the structure's depth is needed. 
For the streamer, a cylindrical geometry appears suitable but, in contrast, the envelope cannot be well represented by such a geometry, as demonstrated in the channel maps of \ce{HC3N}, \ce{C2H}, and c-\ce{C3H2} (see Appendix A of \citetalias{tanious2024}). Therefore, the use of the projected envelope width on the sky cannot be used as a proxy for the envelope depth. A dedicated study of the envelope geometry is therefore needed, but requires higher spectral and velocity resolution observations to probe its 3D shape. As a first guess based on the c-\ce{C3H2} channel map (see Fig.~A.7 of \citetalias{tanious2024}), most of the envelope emission is seen on two channels (6.61 and 6.84 km~s$^{-1}$), while the streamer emission is seen on three channels (6.84, 7.07, and 7.29 km~s$^{-1}$). Assuming that velocity is a proxy of depth, the envelope would have a depth that is about two-thirds that of the streamer. Using the ortho-c-\ce{C3H2} methodology for the envelope (excluding the streamer), we derived an envelope mass $\sim0.030\,M_\odot$ (see Sect.~\ref{sec:env-mass-estimate}). Including the streamer, the total envelope mass is $\sim 0.048 \, M_\odot$. Thus, we can confirm that a significant part of the total envelope mass is contained in the streamer (i.e., $\geq 38$\%). We note that the derived total mass agrees within a factor of $\sim2$ with the above values reported in the literature and obtained from independent methods. We emphasize, however, that this estimate should be considered as an order of magnitude determination, as it remains sensitive to poor constraints on the line-of-sight depth and molecular abundance. A more reliable assessment of the envelope mass would require a dedicated study, for instance, combining multi-molecule non-LTE radiative transfer modeling with higher-velocity-resolution interferometric data to better resolve the envelope structure. 

In the Class 0 object \object{IRAS 16544-1604}, an opposite conclusion was found, that is, less than 10\% of the protostellar envelope is encompassed in the streamers around this object \citep{kido2025}. This would suggest that most of the protostellar mass at early stages (Class 0) is accreted rapidly on the central object and the disk, leaving streamers as the last mechanism for accretion of the remaining envelope mass at later stages (> Class~I). Indeed, the streamer mass in L1489~IRS is significantly lower than the protostar mass, which suggests the protostar has likely reached its final mass. A similar conclusion was reported after deriving a total mass of 4-7~$\times\,10^{-3}\,M_\odot$ for the two other infalls observed in this source and for which the infall rate was estimated to 4-7~$\times\,10^{-7}\,M_\odot$~yr$^{-1}$ \citep{yen2014}. The derived infall rate of the present streamer is of the same magnitude $\gtrsim 10^{-7}$~$M_\odot$~yr$^{-1}$, which is comparable to the star accretion rate based on the star bolometric luminosity \citep[$\sim \num{2e-7}\,M_\odot$~yr$^{-1}$,][]{prato2009,yen2014}. This result suggests that most of the mass provided by the streamer is accreted by the star. 

Recent numerical simulations suggest that the disk mass in Class 0/I protostars is regulated by the gravitational instability \citep[e.g.,][]{mauxion2024}. In this scenario, if a streamer accretes mass onto the disk, it may become gravitationally unstable and part of its material will accrete onto the star until the disk becomes stable again. This results in mass constantly transiting through the disk, which is likely to have a strong impact on its evolution. Assuming a constant streamer mass over time and the modeled 24.1~kyr infalling time derived in \citetalias{tanious2024} \citep[comparable to the protostellar age of $\sim28$~kyr,][]{sai2022}, the system would have accreted a mass of $\geq 65-257$\% relatively to its current disk's mass (see Sect.~\ref{sec:discussion-streamer-properties}). As the typical lifetime of Class I sources is estimated to be $\sim10^5$~yrs \citep[see references in][]{sai2022}, the same amount of mass is likely to be provided in the future, as the Class I stage would last over the infalling time. This supports the hypothesis made in \citetalias{tanious2024}, namely, that the streamer is at the origin of the external warped disk in this system. Such phenomena are indeed seen in numerical simulations, where the star-disk axis is deeply affected by streamers \citep{kuffmeier2021,kuffmeier2024}, and this has also been  suggested by observations \citep[e.g.,][]{tang2012,ginski2021}. 

Moreover, this assumes that the disk would be (at least partially and may be multiple times) replenished by material coming from the streamer onto the disk. Similarly, the disk in the Class I object \object{VLA~4A} may be replenished by the streamer in less than 40~kyr \citep{hsieh2023}. Also, the properties of the streamer in the present study are consistent with another streamer detected toward another Class I protostar, \object{Per-emb-50} in Perseus, having a star mass similar to L1489~IRS, and with a streamer mass of \num{1.2e-2}~$M_\odot$ and an infall rate of $\sim$~\num{1.3e-6}~$M_\odot$~yr$^{-1}$ \citep{valdivia-mena2022}. Therefore, a similar replenishment would be possible in the \object{Per-emb-50} system. In IRAS~16544-1604, the total mass of the streamers around the source is estimated to be $\num{7.5e-3}\,M_\odot$ and the disk emission is reproduced with a radiative transfer model using a disk mass of $\num{1.0e-2}\,M_\odot$ \citep{kido2025}. This would result in a 75\% contribution of the streamer to the disk mass. However, as discussed in \cite{kido2025}, as the protostar likely did not reach its final mass, all the material in the disk+streamer would be accreted by the protostar before the typical lifetime of Class 0 objects if there is no other external reservoirs feeding the disk and/or the streamers. They conclude that streamers are not the main suppliers of material to the disk for the latter to survive the star formation process. In the L1489~IRS system, the streamer is likely fed by the nearby protostellar core as shown by mosaic observations (see Fig.~\ref{fig:snr-maps-simplified}), ensuring mass provision to the streamer. Moreover, as most of the remaining envelope mass is contained in the streamer, it is thus acting as the dominant process of accretion in this Class I object.

Streamers may have strong impacts on the chemistry. For instance, a super-deuteration has been reported in the Class I system \object{IRS~63}, where the shock at the landing point of the streamer onto the disk released \ce{D2CO} presumably formed on grain mantles into the gas phase, likely affecting the disk chemistry \citep{podio2024}. As accretion shocks were also suggested in L1489~IRS \citep[\citetalias{tanious2024}]{yen2014, yamato2023}, a similar behavior may be found in this system. Moreover, the streamer acts as a gas bridge between the core and the YSO. Therefore, the chemical history and evolution of the protoplanetary disk should be intimately linked to the core through the streamer. Deuterated species, which are preferentially formed in low temperature conditions \citep[e.g., in dense cores,][]{ceccarelli2014}, might travel through the streamer to finally end up in the disk. As the D/H ratio is one of the most commonly used probes to trace inheritance versus chemical reset, it seems crucial to investigate the spatial evolution of this ratio from the core to the disk through the streamer. More generally, physico-chemical processes are likely to affect molecules that will end in the disk and affect its chemistry. Therefore, understanding the chemical evolution of the disk requires understanding the chemical evolution of the streamer. L1489~IRS appears to serve as an ideal target to investigate this question, as scales from core to disk are accessible in the same system. To dig into this question, access to new molecular survey observations will be necessary.

%%%%%%%%%%%%%%%%%%%%%%%%%%%%%%%%%%%%%%%%%%%%%%%%%%%%%%%%%%%

\section{Conclusions}
\label{sec:conclusions}

We present new NOEMA and IRAM-30m observations toward L1489~IRS in \ce{C2H}, c-\ce{C3H2}, and \ce{HC3N}, identified in \citetalias{tanious2024} as molecular tracers of the longest streamer seen in carbon chain emission in this source. Our main conclusions are summarized below:
\begin{enumerate}
    \item We confirm the connection between the protostellar core L1489 and the YSO L1489~IRS through the streamer, as suggested in \citetalias{tanious2024}. A detailed study of the mosaic used to confirm this connection will be the focus of a forthcoming paper.
    
    \item Using LTE and self-consistent MCMC \texttt{RADEX} non-LTE radiative transfer modeling, we derived \ce{C2H}, c-\ce{C3H2}, and \ce{HC3N} column densities and abundances in the streamer. The derived values are close to those in similar objects. 
    The relatively high \ce{C2H}/c-\ce{C3H2} ratio, with respect to classical core values, possibly points to a region strongly influenced by UV irradiation. However, the comparable \ce{HC3N}/c-\ce{C3H2} ratio between the streamer and the core suggests that the observed chemistry may instead be (or also be) inherited from the dense core. Taken together, these findings indicate that the streamer likely retains a chemical fingerprint from its parental core, rather than being solely shaped by local UV-driven processes. These are the first hints of interstellar heritage in a Class I YSO, mediated by a streamer connecting the system to its natal environment.

    \item We derived lower limits on streamer properties, namely, its mass $\geq (4.67-18.3)\, \times 10^{-3}$~$M_\odot$ and its infalling rate $\geq (1.94-7.57)\, \times 10^{-7}$~$M_\odot$~yr$^{-1}$, with the ranges corresponding to the different molecular tracers. The streamer likely contains a significant part of the remaining envelope mass, suggesting it is the dominant and last accretion mechanism in this Class I object; however, a dedicated study to assess the envelope geometry is required to confirm this result.
    The relative mass provided by the streamer to the disk would be $\geq 65-257$\%, using a disk mass of $M_\text{disk}=\num{7.1e-3}\,M_\odot$ \citep{sai2020}. This  supports a streamer origin for the external warped disk in this system, as well as a fully replenished disk, with strong consequences for the chemistry. 
\end{enumerate}
To obtain better constraints on the \ce{H2} density, more lines (especially non-LTE lines) ought to be included to the MCMC \texttt{RADEX} model to improve the estimation. This would enable a better constraint on streamer properties and, thus, a better evaluation of the mass provided by the streamer to the system. As streamers are thought to have a strong impact on  disks chemistry, 
detailed studies from cores to disks through streamers need to be conducted. L1489~IRS is an ideal target for addressing this question, as it will ultimately offer some hints on interstellar heritage from very early stages to more evolved stages of star formation, using a single system connecting all scales.

%%%%%%%%%%%%%%%%%%%%%%%%%%%%%%%%%%%%%%%%%%%%%%%%%%%%%%%%%%%

\begin{acknowledgements}
  The authors thank the anonymous referee for the interest and valuable comments which helped to improve the paper.
  The authors also thank the IRAM staff for their invaluable work making these observations possible. 
  M.T., R.L.G., and A.F. are also grateful to Jérôme Pety for his advices for the data reduction.
  This work was supported by the Programme National “Physique et Chimie du Milieu Interstellaire” (PCMI) of CNRS/INSU with INC/INP co-funded by CEA and CNES. 
  This study is based on observations carried out under project numbers 184-20, S20AH and W20AJ (PI: Le Gal), and 097-24, S24AR (PIs: Tanious \& Le Gal), with IRAM-30m and IRAM Interferometer NOEMA. IRAM is supported by INSU/CNRS (France), MPG (Germany) and IGN (Spain). 
  This research made use of spectroscopic and collisional data from the EMAA database \citep{emaa}. EMAA is supported by the Observatoire des Sciences de l’Univers de Grenoble (OSUG).
  This research made use of \texttt{astropy} \citep{astropy2013,astropy2018,astropy2022}, \texttt{corner} \citep{corner}, \texttt{emcee} \citep{emcee}, \texttt{GILDAS} \citep{gildas}, \texttt{gofish} \citep{gofish}, \texttt{matplotlib} \citep{matplotlib2007}, \texttt{numpy} \citep{numpy2020}, \texttt{RADEX} \citep{vandertak2007}, \texttt{TIPSY} \citep{tipsy2024}, and \texttt{ultraplot} \citep{proplot}.
\end{acknowledgements}

%%%%%%%%%%%%% BIBLIOGRAPHY %%%%%%%%%%%%%%%%%
% \bibliographystyle{aa}
\bibliographystyle{aa}
\bibliography{refs}

% \listofobjects

%%%%%%%%%%%%%%% APPENDIX %%%%%%%%%%%%%%%%%%%
\begin{appendix}

\onecolumn
\section{Line emission maps}

\begin{figure*}[!h]
    \centering
    \includegraphics[width=\linewidth]{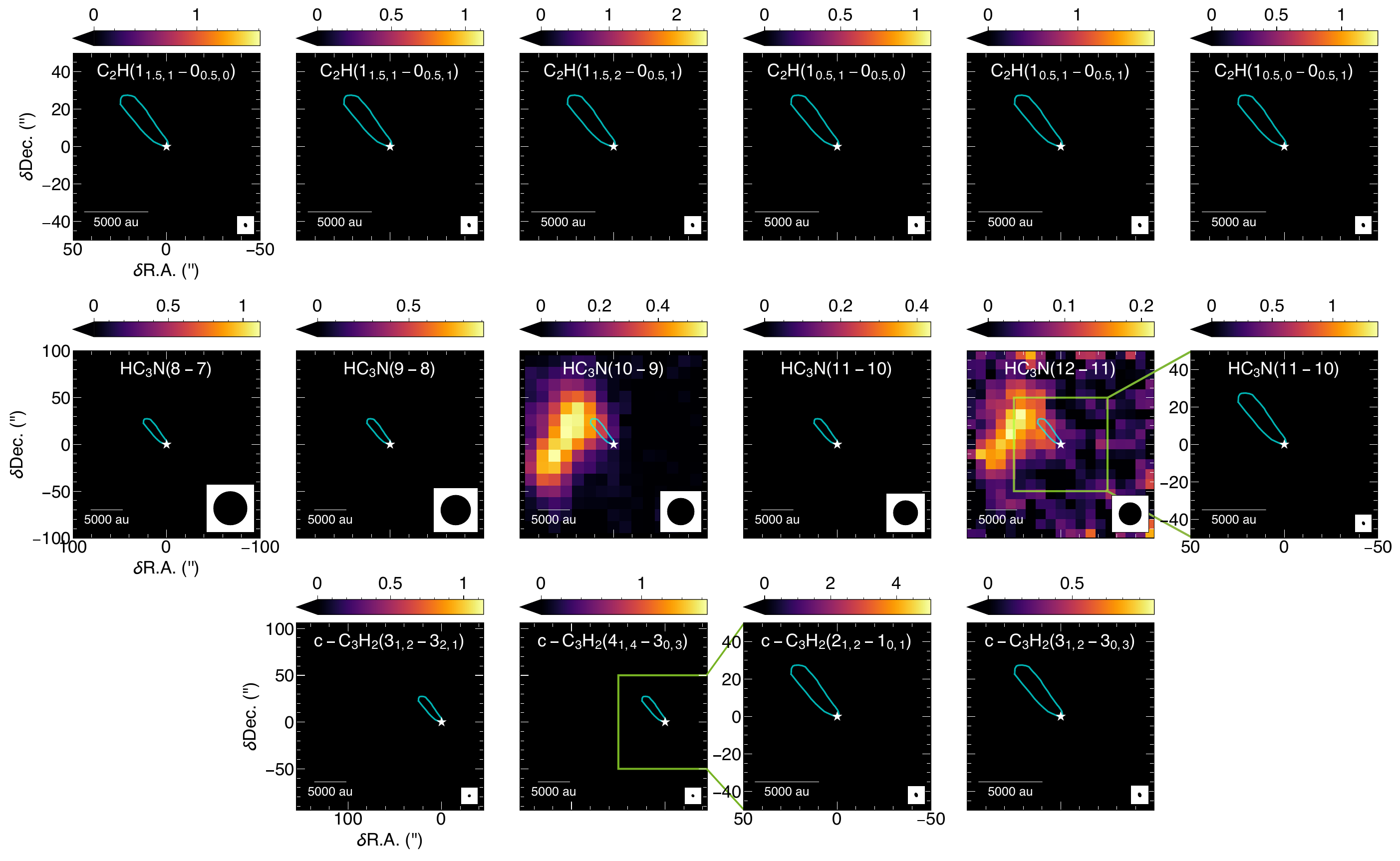}
    \caption{
    Gallery of integrated intensity maps (in K~km~s$^{-1}$) of observed lines used in this work. The line name is indicated on the top of each panel while their beam is shown in the lower right corner. The star indicates the position of L1489~IRS. The cyan contour corresponds to the targeted emission for the analysis (see Sect.~\ref{sec:data-preprocessing}). \textbf{Top row:} Single-field combined (IRAM-30m + NOEMA) observations of \ce{C2H}. \textbf{Middle row:} IRAM-30m (columns 1 to 5) and combined (column 6) observations of \ce{HC3N}. \textbf{Bottom row:} Mosaic (first and second column) and single-field (third and fourth column) combined observations of \ce{c-C3H2}.
    }
    \label{fig:mom0-maps}
\end{figure*}

\begin{figure*}[!h]
    \centering
    \includegraphics[width=\textwidth]{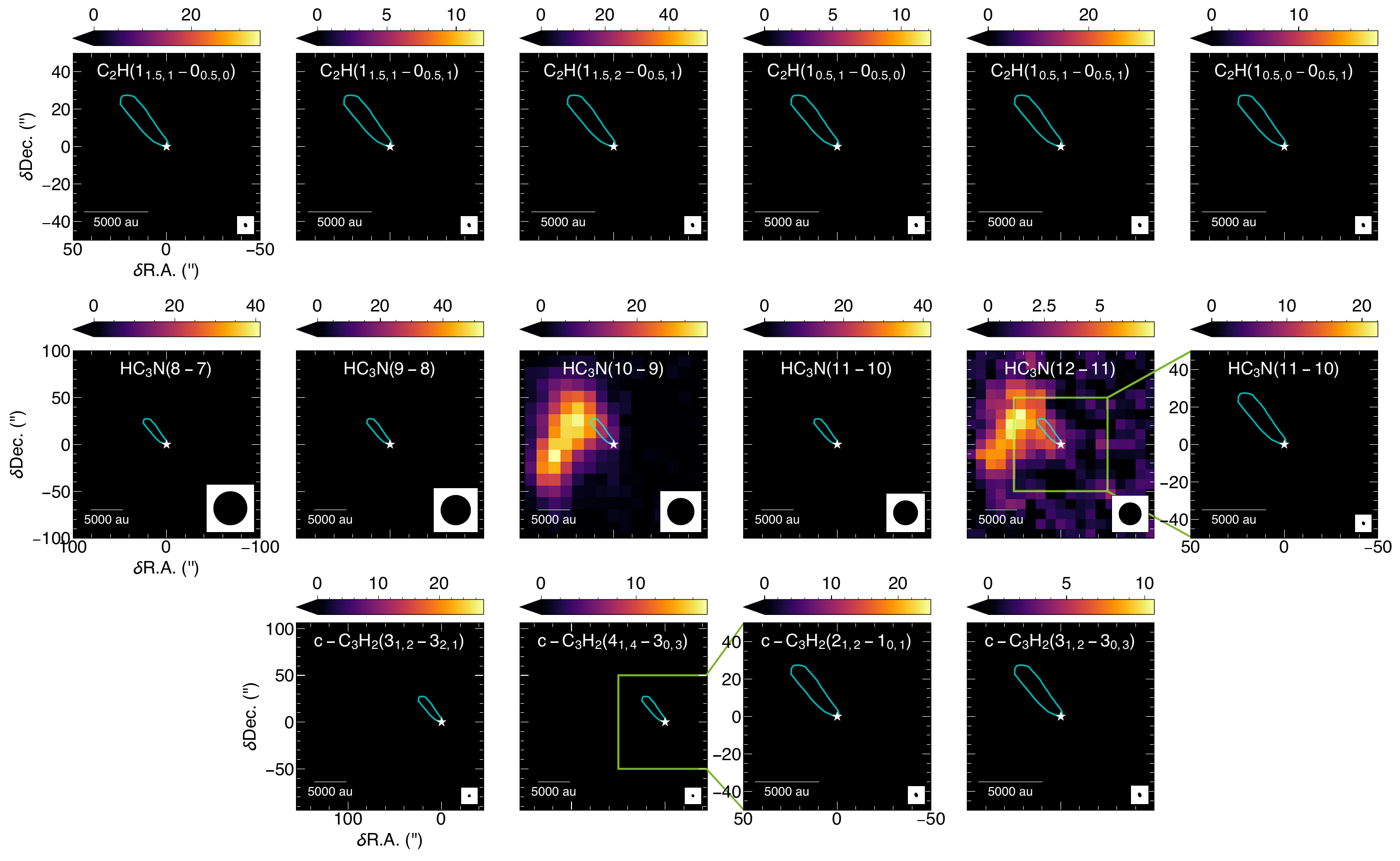}
    \caption{
    Same as Fig.~\ref{fig:mom0-maps}, but for the S/N maps of observed lines used in this work.
    }
    \label{fig:snr-maps}
\end{figure*}

Figs.~\ref{fig:mom0-maps} and \ref{fig:snr-maps} show the integrated intensity and S/N maps for each line used in this work, produced according to the procedure described in Sect.~\ref{sec:data-preprocessing}.

\FloatBarrier

\twocolumn
\newpage
\section{HFS Fit parameters}

\begin{figure}[!h]
    \centering
    \includegraphics[width=0.6\linewidth]{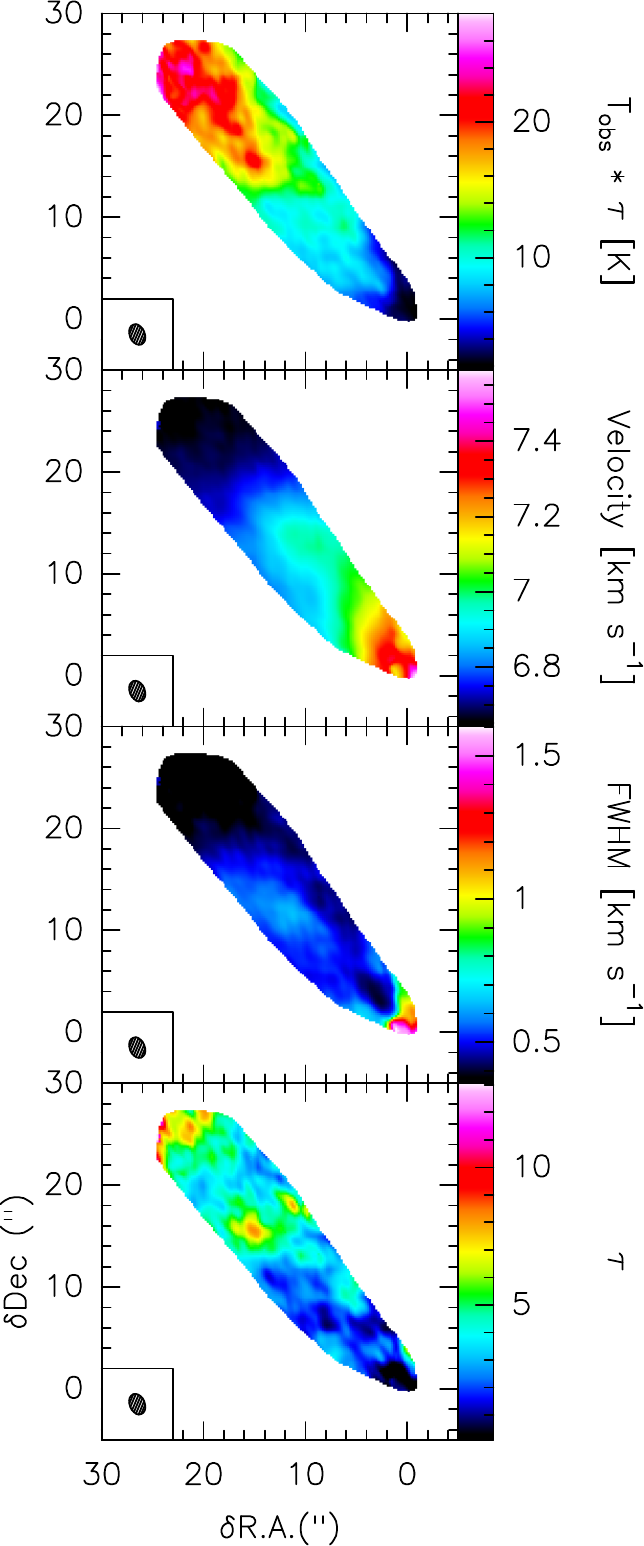}
    \caption{Maps of parameters obtained after the HFS fitting of \ce{C2H}. \textbf{From top to bottom:} $T_\text{obs}\times\tau$ in K, velocity of the main component in km~s$^{-1}$, FWHM of the main component in km~s$^{-1}$, and group total opacity, $\tau$.}
    \label{fig:c2h-hfs-results}
\end{figure}

Figure~\ref{fig:c2h-hfs-results} shows the maps of the derived parameters from the HFS fitting of \ce{C2H} described in Sect.~\ref{sec:hfs} and \ref{sec:results-c2h}.

\FloatBarrier

\section{Critical densities}
\label{sec:critical-densities}

The critical density for a particular level, $i$, at a given kinetic temperature, $\Tkin$, is defined in the optically thin limit as
\begin{equation}
    n^{cr}_{i}(\Tkin) = \frac{\sum_{j<i} A_{ij}}{\sum_{j\ne i} k_{ji}(\Tkin)}
,\end{equation}
where $A_{ij}$ is the Einstein $A$ coefficient and the sum runs over all possible emissions, and $k_{ij}(\Tkin)$ is the rate coefficient for collisional de-population of level $i$, summed over all possible excitations and de-excitations. The level $i$ can be thermalized at the kinetic temperature, i.e., populated at LTE only at densities much larger (by typically one or two orders of magnitude) than the critical one. Conversely, at densities well below the critical density, collisions can be neglected and the level $i$ is in radiative equilibrium.

In practice, for the upper levels listed in Table~\ref{table:critical-densities}, critical densities at $\Tkin=5$, 10 and 20~K are found to range between $\sim 2\times 10^4$ and $\sim 5\times 10^5$~cm$^{-3}$ for the molecular transitions used in this work, meaning that these levels are thermalized at densities larger than $n({\rm H_2})\sim 10^6$~cm$^{-3}$. As a result, the selected molecular transitions (i.e., those of \ce{C2H}, ortho-c-\ce{C3H2}, and \ce{HC3N}) should be good probes of density in regions where $10^3 \lesssim n({\rm H_2})\lesssim  10^6$~cm$^{-3}$.

\begin{table}
\caption{Critical densities for the upper levels involved in the observed transitions of C$_2$H, ortho-c-C$_3$H$_2$ and HC$_3$N. Critical densities of HCN were added for discussion purposes (see Sect.~\ref{sec:discussion-h2-density}).}             
\label{table:critical-densities}      
\centering                          
\renewcommand{\arraystretch}{1.5} 
\small
\resizebox{\linewidth}{!}{
\begin{tabular}{l c c  c c}        
\hline\hline                 
    Species & Upper level & \multicolumn{3}{c} {Critical density (cm$^{-3}$)}  \\
    & & 5~K & 10~K & 20~K \\

\hline    
    \ce{C2H} & $1_{1.5,1}$  & -- & \num{3.46e4} & \num{1.97e4} \\
    
             & $1_{1.5,2}$  & -- & \num{3.73e4} & \num{2.07e4} \\
             
             & $1_{0.5,1}$  & -- & \num{2.86e4} & \num{1.74e4} \\
             
             & $1_{0.5,0}$  & -- & \num{2.55e4} & \num{1.64e4} \\

\hline  
    ortho-\ce{c}-\ce{C3H2} & $2_{1,2}$ &  \num{8.90e4} & \num{6.74e4} & \num{5.46e4} \\
    
                           & $3_{1,2}$ &  \num{2.17e5} & \num{1.87e5} & \num{1.66e5} \\
                           
                           & $4_{1,4}$ &  \num{4.82e5} & \num{4.15e5} & \num{3.66e5} \\

\hline 
    \ce{HC3N} & $8$   & \num{6.55e4} & \num{5.64e4} & \num{4.53e4} \\
              & $9$   & \num{9.20e4} & \num{8.06e4} & \num{6.51e4} \\
              & $10$  & \num{1.26e5} & \num{1.11e5} & \num{9.01e4} \\
              & $11$  & \num{1.63e5} & \num{1.47e5} & \num{1.20e5} \\
              & $12$  & \num{2.11e5} & \num{1.91e5} & \num{1.57e5} \\
\hline 
    \ce{HCN}  & $1$   & \num{6.35e5} & \num{4.81e5} & \num{2.71e5} \\
              & $2$   & \num{6.20e6} & \num{4.31e6} & \num{2.30e6} \\
              & $3$   & \num{1.83e7} & \num{1.27e7} & \num{7.14e6} \\
              & $4$   & \num{3.35e7} & \num{2.57e7} & \num{1.60e7} \\
\hline                                   
\end{tabular}
}
\tablefoot{Spectroscopic and collisional data are taken respectively from the CDMS \citep{muller2005} and EMAA databases. For \ce{C2H}, no collisional rates were available at 5~K.
}
\tablebib{Collisional data for \ce{C2H} \citep{Pirlot2023}, c-\ce{C3H2} \citep{BenKhalifa2019}, \ce{HC3N} \citep{Hily2018}, and HCN \citep{Hernandez2017}.}
\end{table}

\FloatBarrier

\newpage
\section{MCMC \texttt{RADEX} results}
\label{sec:mcmc-results-appendix}
We present in this section the corner plots, the reproduced integrated intensities, and the excitation temperatures from MCMC \texttt{RADEX} explorations in this work: for \ce{C2H} in Sect.~\ref{sec:mcmc-results-c2h}, ortho-c-\ce{C3H2} in Sect.~\ref{sec:mcmc-results-occ3h2}, and \ce{HC3N} in Sect.~\ref{sec:mcmc-results-hc3n}. The best \texttt{RADEX} model in the following corresponds to the model ran with median parameters from the MCMC exploration.

\subsection{\ce{C2H}}
\label{sec:mcmc-results-c2h}

\begin{table}[!h]
    \caption{Observed and MCMC \texttt{RADEX} best model integrated intensities and excitation temperature for \ce{C2H}.}             
    \label{table:c2h-obs_vs_RADEX}      
    \centering                          
    \renewcommand{\arraystretch}{1.5} 
    \begin{tabular}{c c c c}        
    \hline\hline      
                   &          & \multicolumn{2}{c}{Best \texttt{RADEX} model} \\
        Transition & Observed & Int. intensity & $\Tex$ \\
        & (K~km~s$^{-1}$) & (K~km~s$^{-1}$) & (K)\\  
    \hline    
        $1_{1.5,1}\to0_{0.5,0}$  & $1.003\pm0.115$ & 1.115 & 6.4 \\
        $1_{1.5,1}\to0_{0.5,1}$  & $0.351\pm0.074$ & 0.327 & 6.6 \\
        $1_{1.5,2}\to0_{0.5,1}$  & $1.608\pm0.171$ & 1.621 & 6.6 \\
        $1_{0.5,1}\to0_{0.5,0}$  & $0.323\pm0.066$ & 0.326 & 6.5 \\
        $1_{0.5,1}\to0_{0.5,1}$  & $1.081\pm0.124$ & 1.154 & 6.6 \\
        $1_{0.5,0}\to0_{0.5,1}$  & $0.584\pm0.085$ & 0.601 & 6.7 \\
    \hline                                   
    \end{tabular}
\end{table}

\begin{figure}[!h]
    \centering
    \includegraphics[width=\linewidth]{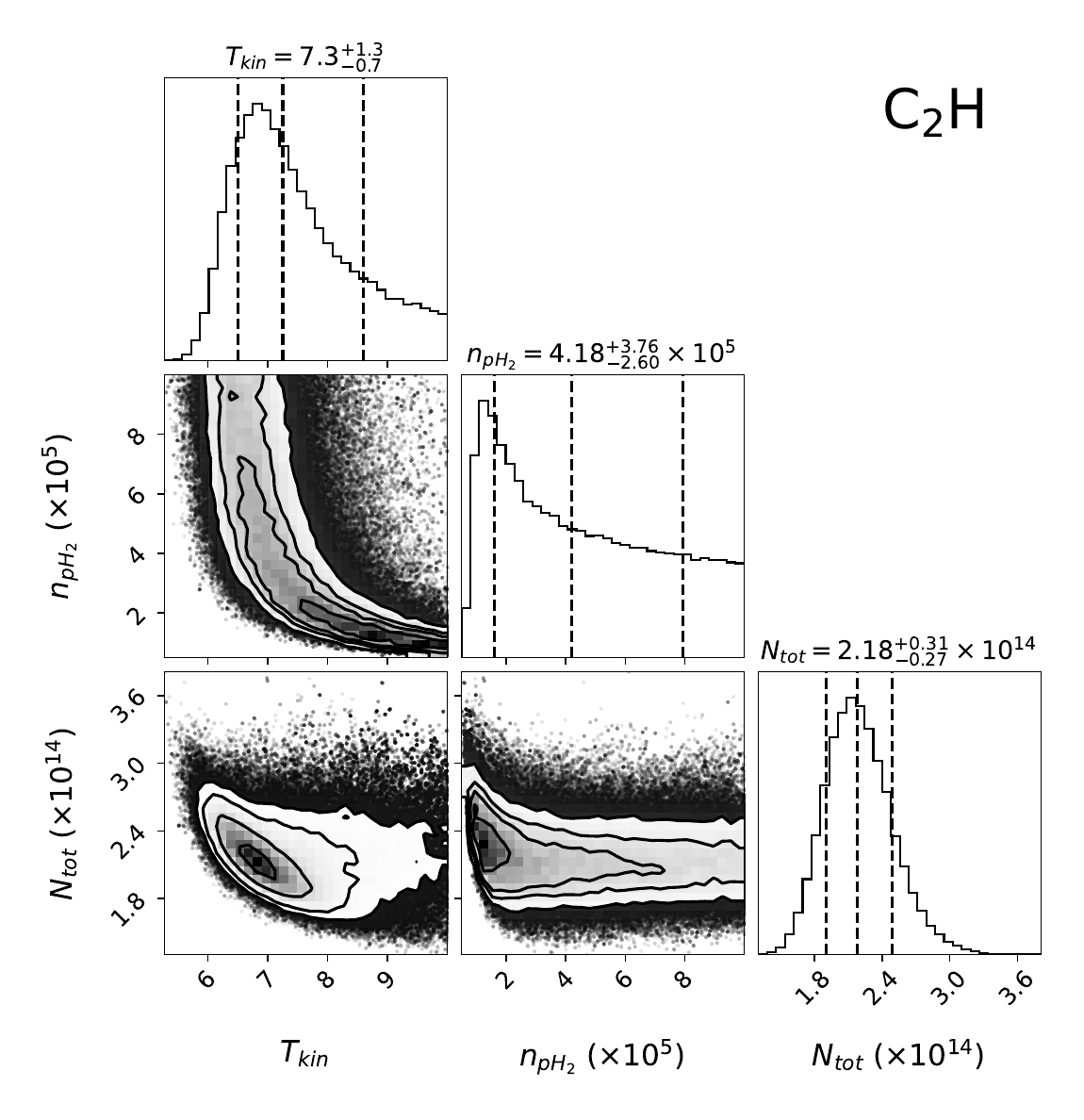}
    \caption{Corner plot obtained from the MCMC \texttt{RADEX} exploration for \ce{C2H}. The three dashed lines on histogram panels correspond to the 16\textsuperscript{th}, 50\textsuperscript{th} and 84\textsuperscript{th} percentiles of the distribution. $\Tkin$ is in K, \npH~in cm$^{-3}$, and $\Ntot$ in cm$^{-2}$.
    }
    \label{fig:corner-plot-c2h}
\end{figure}

\FloatBarrier
\newpage
\subsection{ortho-c-\ce{C3H2}}
\label{sec:mcmc-results-occ3h2}
\begin{table}[!h]
    \caption{Same as Table~\ref{table:c2h-obs_vs_RADEX}, but for c-\ce{C3H2}.}             
    \label{table:cc3h2-obs_vs_RADEX_allres}      
    \centering                          
    \renewcommand{\arraystretch}{1.5} % Default value: 1, space between rows
    \begin{tabular}{c c c c}        
    \hline\hline                 
                   &          & \multicolumn{2}{c}{Best \texttt{RADEX} model} \\
        Transition & Observed & Int. intensity & $\Tex$ \\
        & (K~km~s$^{-1}$) & (K~km~s$^{-1}$) & (K)\\   
    \hline    
        $2_{1,2}\to1_{0,1}$  & $0.728\pm0.097$ & 0.714 & 14.3\tablefootmark{*} \\
        $3_{1,2}\to2_{2,1}$  & $0.433\pm0.090$ & 0.349 & 8.6 \\
        $3_{1,2}\to3_{0,3}$  & $0.135\pm0.021$ & 0.154 & 9.4 \\
        $4_{1,4}\to3_{0,3}$  & $0.624\pm0.145$ & 0.590 & 8.2 \\
    \hline                                   
    \end{tabular}
    \tablefoot{
        \tablefoottext{*}{This line has an excitation temperature higher than the kinetic temperature ($\Tkin=12.4$~K, see Table~\ref{table:summary-TR-results}), suggesting a density regime favorable to a population inversion.}
    }
\end{table}

\begin{figure}[!h]
    \centering
    \includegraphics[width=\linewidth]{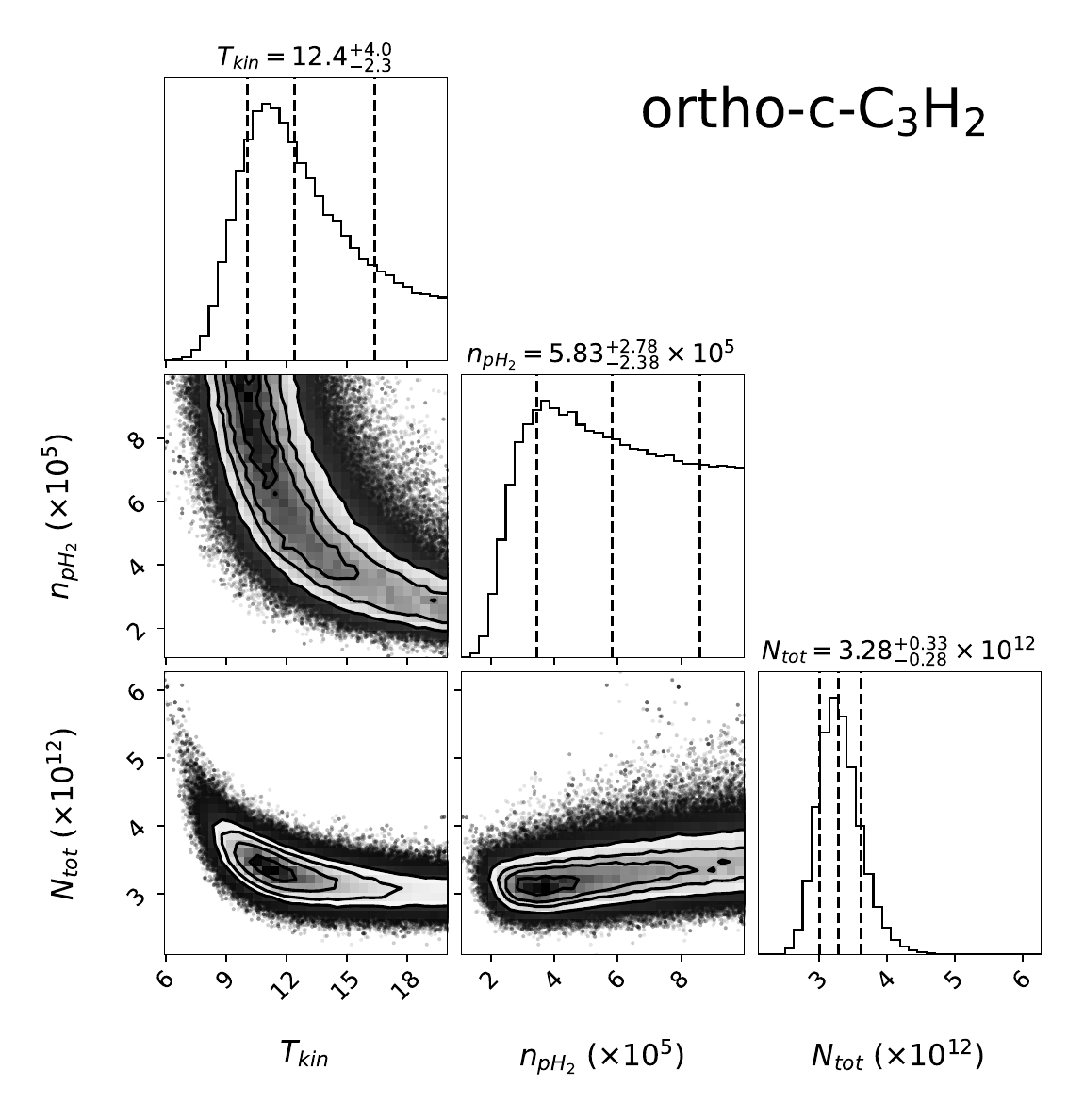}
    \caption{Same as Fig.~\ref{fig:corner-plot-c2h}, but for ortho-c-\ce{C3H2}.
    }
    \label{fig:corner-plot-ocC3H2}
\end{figure}

\FloatBarrier

\newpage
\subsection{\ce{HC3N} on IRAM-30m lines}
\label{sec:mcmc-results-hc3n}
\begin{table}[!h]
    \caption{Same as Table~\ref{table:c2h-obs_vs_RADEX}, but for \ce{HC3N} IRAM-30m lines (corrected by the beam filling factor, $f=0.253$, see Sect.~\ref{sec:results-hc3n}).}             
    \label{table:hc3n-obs_vs_RADEX_allres}      
    \centering                          
    \renewcommand{\arraystretch}{1.5} 
    \begin{tabular}{c c c c}        
    \hline\hline                 
                   &          & \multicolumn{2}{c}{Best \texttt{RADEX} model} \\
        Transition & Observed & Int. intensity & $\Tex$ \\
        & (K~km~s$^{-1}$) & (K~km~s$^{-1}$) & (K)\\   
    \hline    
        $8\to7$   & $1.747\pm0.186$ & 1.637 & 6.7 \\
        $9\to8$   & $1.387\pm0.142$ & 1.373 & 6.6 \\
        $10\to9$  & $0.897\pm0.095$ & 1.033 & 6.4 \\
        $11\to10$ & $0.759\pm0.079$ & 0.683 & 6.3 \\
        $12\to11$ & $0.368\pm0.055$ & 0.389 & 6.1 \\
    \hline                                   
    \end{tabular}
\end{table}

\begin{figure}[!h]
    \centering
    \includegraphics[width=\linewidth]{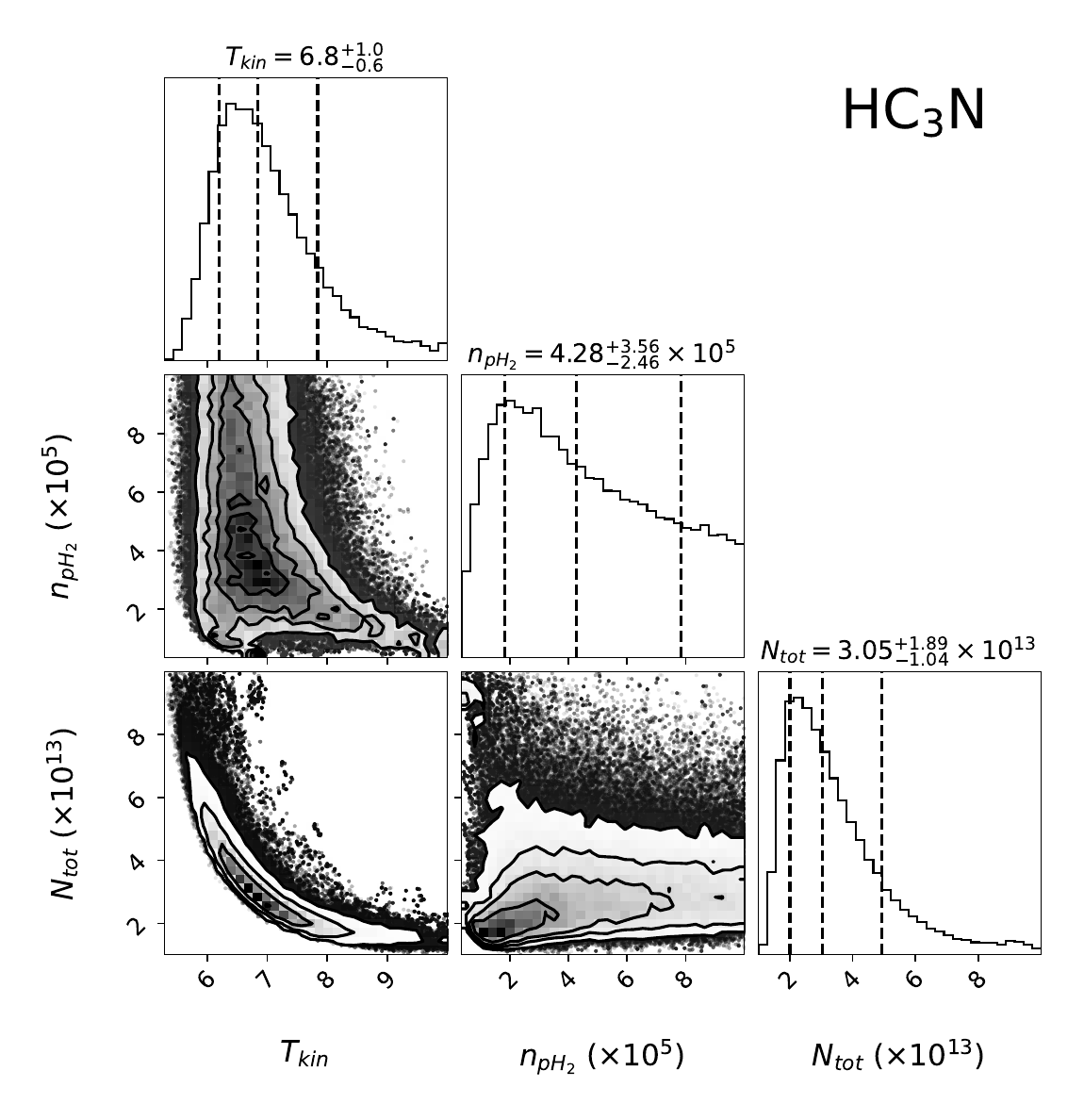}
    \caption{Same as Fig.~\ref{fig:corner-plot-c2h}, but for \ce{HC3N}.
    }
    \label{fig:corner-plot-hc3n}
\end{figure}

\FloatBarrier

\section{c-\ce{C3H2} rotational diagram fit quality}
\begin{figure}[!h]
    \centering
    \includegraphics[width=\linewidth]{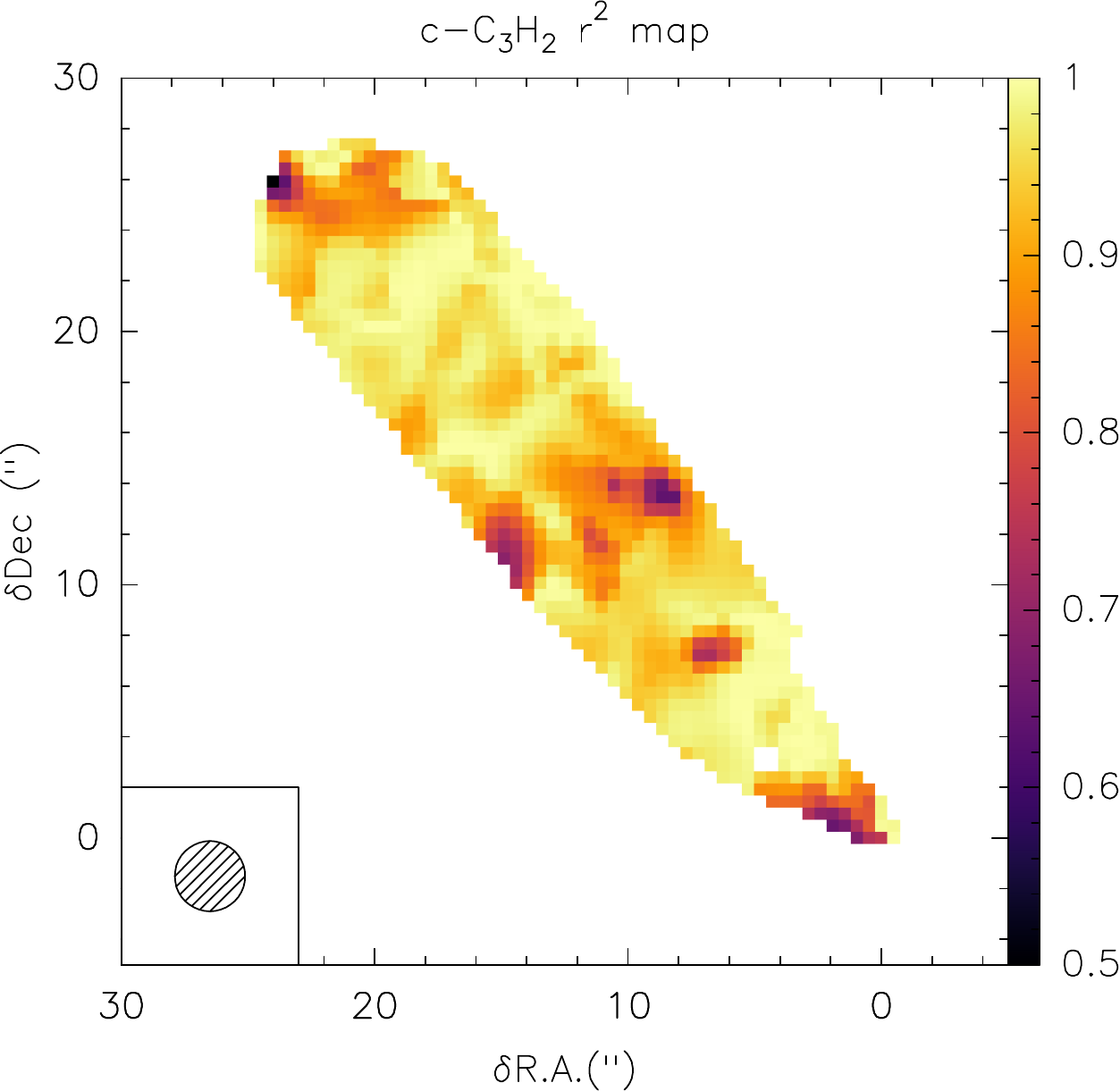}
    \caption{$r^2$ map obtained from the linear fitting in each pixel of the rotational diagram for ortho-c-\ce{C3H2}.}
    \label{fig:rd-ocC3H2-r2}
\end{figure}

Fig.~\ref{fig:rd-ocC3H2-r2} displays the $r^2$ map obtained from the linear fitting in each pixel of the rotational diagram for ortho-c-\ce{C3H2} described in Sect.~\ref{sec:results-ocC3H2}.

\FloatBarrier

\newpage
\section{Envelope mass estimate}
\label{sec:env-mass-estimate}

\begin{figure}[!h]
    \centering
    \includegraphics[width=0.65\linewidth]{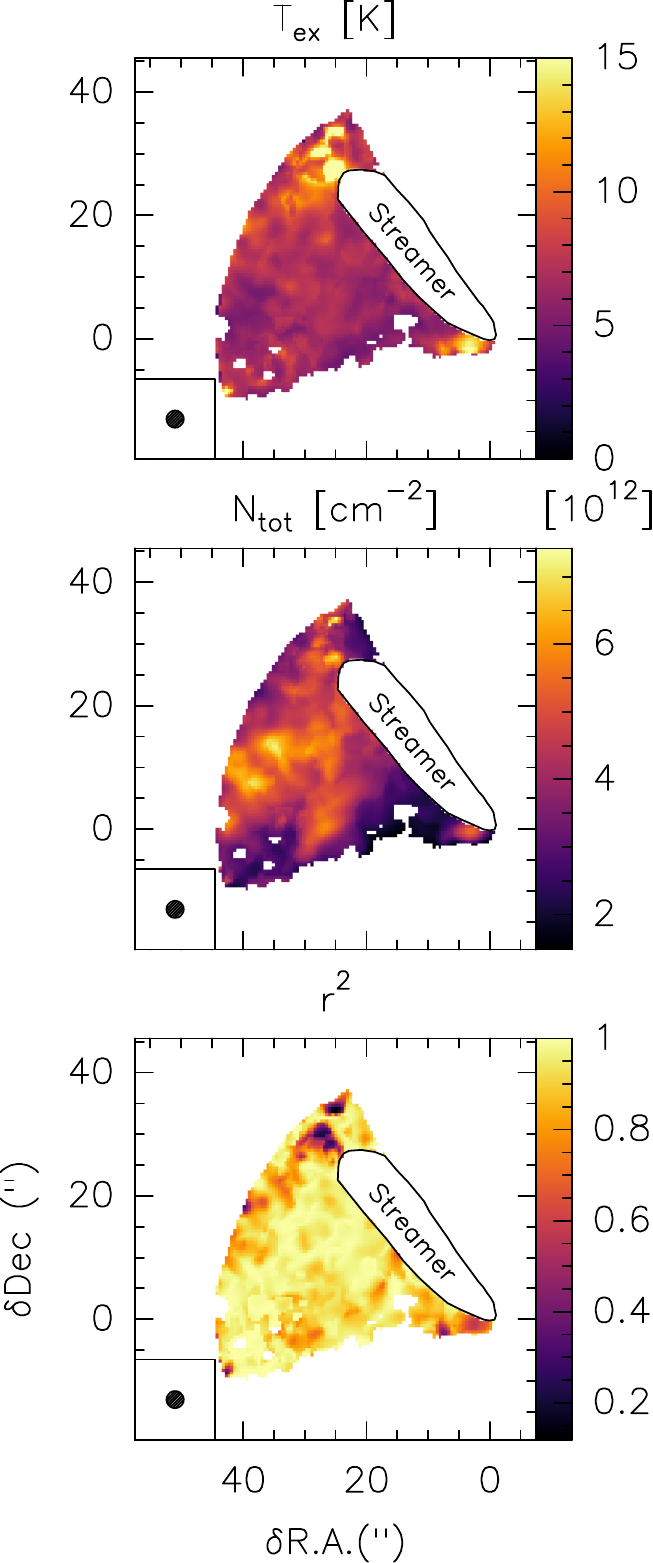}
    \caption{Results of the rotational diagram method applied pixel by pixel on ortho-c-\ce{C3H2} lines for the envelope. The streamer is indicated with the black contour for reference. \textbf{Top:} excitation temperature $\Tex$. \textbf{Middle:} molecular column density $\Ntot$. \textbf{Bottom:} $r^2$ map.}
    \label{fig:RD_cC3H2_env}
\end{figure}

\begin{figure}[!h]
    \centering
    \includegraphics[width=\linewidth]{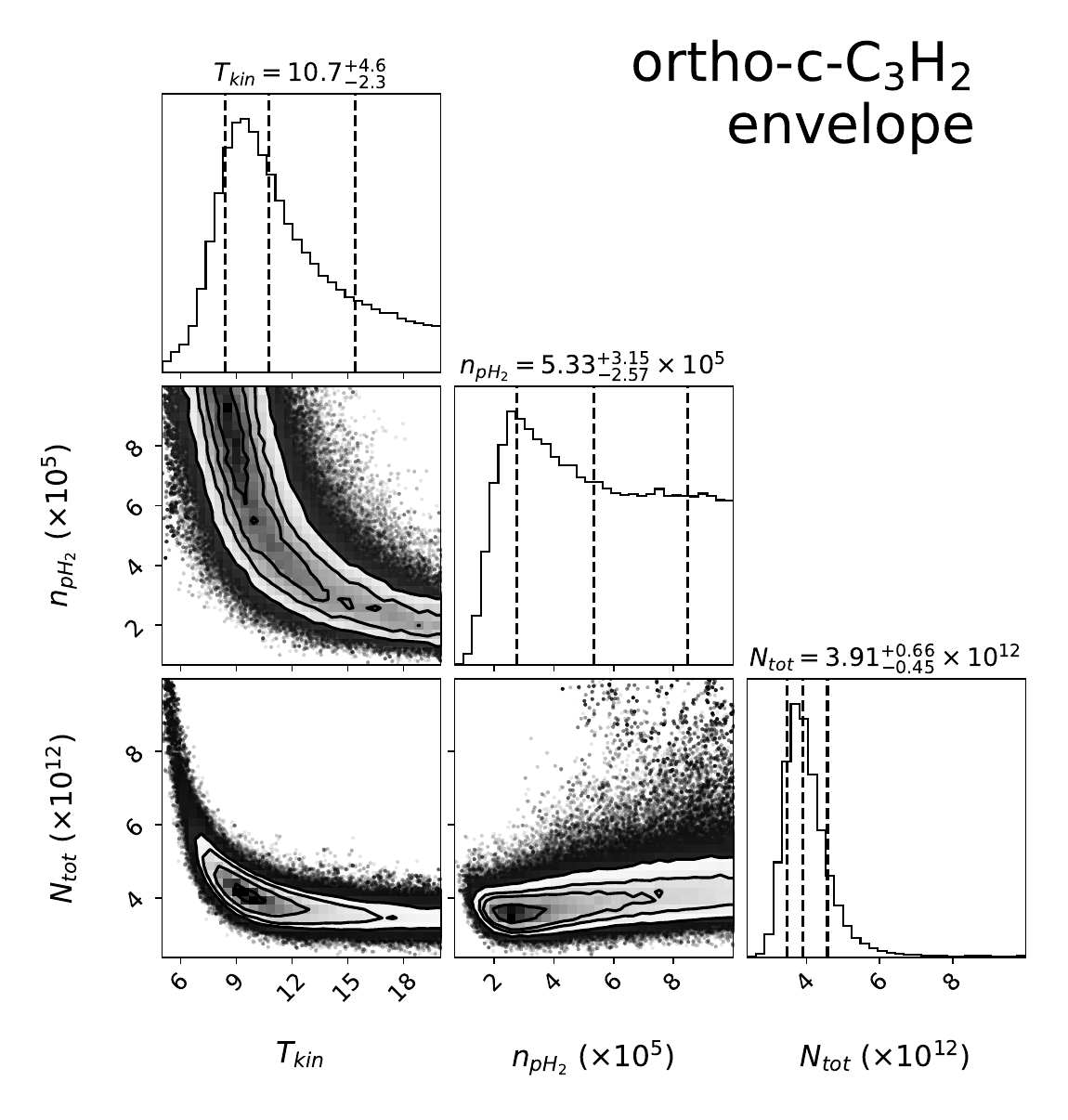}
    \caption{Same as Fig.~\ref{fig:corner-plot-c2h}, but for ortho-c-\ce{C3H2} in the envelope.}
    \label{fig:corner-plot-c3h2-envelope}
\end{figure}

Similarly to the procedure applied to the streamer, we applied the rotational diagram method described in Sect.~\ref{sec:RD} to ortho-c-\ce{C3H2} lines on the envelope excluding the streamer. The results are presented in Fig.~\ref{fig:RD_cC3H2_env}. No substantial variation of the excitation temperature nor the column density is seen over the envelope, with $\Tex=6.9^{\,+1.8}_{\,-1.2}$~K, and $\Ntot=(3.73^{\,+1.57}_{\,-0.91})\,\times 10^{12}$~cm$^{-2}$, which are close to the streamer's values. Using the MCMC \texttt{RADEX} on the observed averaged integrated intensities over the envelope (see Table~\ref{table:cc3h2-obs_vs_RADEX_allres_env}) with initial values of $\Tkin$ and $\Ntot$ set by the rotational diagram, we obtained the corner plot shown in Fig.~\ref{fig:corner-plot-c3h2-envelope} after removing 1500 burn-in steps.
We inferred the kinetic temperature $\Tkin=10.7^{\,+4.6}_{\,-2.3}$~K, another estimation of the column density $\Ntot=(3.91^{\,+0.66}_{\,-0.45})\,\times 10^{12}$~cm$^{-2}$, and a lower limit on the para-\ce{H2} density \npH$\geq \num{2.69e5}$~cm$^{-3}$. 
Using an envelope depth of $2/3$ that of the streamer in Eq.~\eqref{eq:abundance}, we derived an upper limit on the abundance of ortho-c-\ce{C3H2} in the envelope of $\leq \num{1.46e-9}$. This leads to an estimation of the lower limit on the column density of \ce{H2} within the envelope of $\NHtwo \geq \num{1.94e21}$~cm$^{-2}$. Eventually, this gives an envelope mass of $\sim 0.030 \, M_\odot$. Including the streamer, the total envelope mass is $\sim 0.048 \, M_\odot$. This estimate should be interpreted only as an order-of-magnitude value, since it remains sensitive to uncertainties in the line-of-sight depth and molecular abundance. A more reliable determination of the envelope mass would require a dedicated analysis, as discussed in Sect.~\ref{sec:discussion-streamer-impact}.

\begin{table}[!h]
    \caption{Same as Table~\ref{table:c2h-obs_vs_RADEX} for ortho-c-\ce{C3H2}, but for the envelope.}             
    \label{table:cc3h2-obs_vs_RADEX_allres_env}      
    \centering                          
    \renewcommand{\arraystretch}{1.5} 
    \begin{tabular}{c c c c}        
    \hline\hline                 
                   &          & \multicolumn{2}{c}{Best \texttt{RADEX} model} \\
        Transition & Observed & Int. intensity & $\Tex$ \\
        & (K~km~s$^{-1}$) & (K~km~s$^{-1}$) & (K)\\   
    \hline    
        $2_{1,2}\to1_{0,1}$  & $0.866\pm0.120$ & 0.856 & 11.3 \\
        $3_{1,2}\to2_{2,1}$  & $0.385\pm0.089$ & 0.367 & 7.6 \\
        $3_{1,2}\to3_{0,3}$  & $0.146\pm0.029$ & 0.162 & 8.1 \\
        $4_{1,4}\to3_{0,3}$  & $0.596\pm0.142$ & 0.577 & 7.2 \\
    \hline                                   
    \end{tabular}
\end{table}

\end{appendix}

\end{document}